\documentclass[aps,a4,epsf]{revtex4}
\usepackage{graphicx}
\usepackage{longtable}
\usepackage{amsmath}
\usepackage[all]{xy}

\newcommand{\be}{\begin{equation}}
\newcommand{\ee}{\end{equation}}
\newcommand{\ba}{\begin{eqnarray}}
\newcommand{\ea}{\end{eqnarray}}
\newcommand{\baa}{\begin{eqnarray*}}
\newcommand{\eaa}{\end{eqnarray*}}

\begin{document}

\title{ Out-of-equilibrium transport in a typical multi-terminal setup}
\author{In\`es Safi}

\affiliation{Laboratoire de Physique des Solides
 Universit\'e Paris-Sud, 91405 Orsay CEDEX}

\begin{abstract}

We develop a general out-of-equilibrium framework for a typical three-terminal setup of common use: an injector, which can be for instance an interacting wire or an STM electrode, coupled by both extended tunneling and Coulomb interactions to an inhomogeneous wire with any range of interactions and scattering processes, and connected to two reservoirs through interacting or noninteracting leads. Some of the crucial results we obtain are of relevance to other geometries with multiple quantum wires, single wall carbon nanotubes or edge states in the Fractional Quantum Hall Effect (FQHE). We show that the voltage of the injector does not cut the flow of relevant scattering processes in the wire. Either a grounded or a semi-infinite wire at too low temperature is driven into the strong coupling regime and eventually to its fixed point, for instance to tunneling contacts if one starts from almost ohmic ones.

  We show that the injector induce invasive effects. They are due to non-local backscattering processes generated both by virtual higher order tunneling processes and by Coulomb interactions with the injector. The latter induce in addition screening of interactions in the wire.

For an STM, those effects can drastically mask the probed density of states (DOS).
In the limit of zero temperature, a long and grounded wire is driven to its fixed point where it is disconnected at the tunneling point. Thus instead of the bulk expected DOS, the STM probes the end one.

We analyze current auto- and cross-correlations. We show that the search for the signature of quantum statistics requires to have high enough voltage across the wire and not only in the injector. We also show that the cross-correlations are dominated by their value in the two-terminal geometry. As these are opposite to the current auto-correlations, they are always negative for local scattering processes. This is consistent with the fact that electrons obeying Fermi
 statistics are injected. In particular, this solves a previous paradox as they were found positive in an infinite grounded wire described by the Tomonaga-Luttinger Liquid (TLL) model and coupled by local weak tunneling to a sharp STM.
We give novel scaling laws for current and noise in the presence of a sharp injector, to all orders with respect to a local backscattering, which illustrates our results.
\end{abstract}


\maketitle

\vskip2pc
\narrowtext
\maketitle

\section{Introduction}
Multi-terminal mesoscopic structures have been the subject of an intensive theoretical and experimental activity, for instance in Mach-Zender interferometers or Hanbury-Brown and Twiss (HBT)type setups \cite{henny_hbt,oliver_hbt}.
 The structures can either imply similar systems in the different arms, such as edge states in the Hall effect. They can as well be hybrid, for instance formed by a superconductor and other normal metallic arms. One important category of those hybrid structures is that used in the context of investigation of either a DOS of the system by the scanning tunneling microscopes (STM) or its local potential through atomic force microscopes (AFM). In addition, out-of-equilibrium situations offer a rich behavior, such as a superconductor tip coupled to a diffusive wire \cite{pothier_tunnel}.

Many of the non-local mesoscopic transport features have been treated with success by the scattering approach. This is not suited however to treat for instance Coulomb blockade or Kondo physics \cite{kondo_three,shah_kondo}, and more generally to include systematically Coulomb interactions either in the arms or at the junctions. Some progress in this direction was made through a self-consistent determination of the electric potential. Nevertheless, this is not appropriate to treat typical systems where such approximation is not valid, even when interactions are weak. This is  precisely the case for interacting one-dimensional systems. They are described in some limits by the Tomonaga-Luttinger liquid (TLL), characterized by a parameter $g$ roughly related to the interactions by $g\simeq (1+U/2\pi v_F)^{-1}$. Even though the RPA is sufficient for the two-terminal pure wire \cite{blanter,ines_ann,ines_epj}, it cannot handle the presence of additional scattering processes, even for weak $U$, neither many-terminal geometries.

Both experimental and theoretical studies of junctions with many TLL arms have emerged recently, where more rich behavior raises compared to the two-terminal case. Those are realized with edge states in the Fractional Quantum Hall Effect (FQHE), for instance through Mach-Zender interferometry or in shot-noise investigation of dilute quasi particles \cite{heiblum_dilute,kane_fisher_dilute}. They have been achieved as well through crossed \cite{egger_cross,gao_crossed}  Y-type junctions of single wall carbon nanotubes \cite{papa_y,satishkumar_y,li_y,sen_1,trauz_y} or Coulomb drag physics\cite{flensberg,nazarov_drag,egger_drag,drag_review,drag_exp}. There have been theoretical predictions for the multi-lead point-contact tunneling \cite{nayak} and HBT setups in the FQHE \cite{ines_prl,vish_hall,kane_telegraph,kim_hbt}.

The category of multi-terminal geometries based on STM or AFM is the most  frequently used experimentally\cite{bockrath_99,bachtold_afm,gao_four}, but not many theoretical works have been devoted to its  investigation \cite{lee_stm,eggert_dos,pham_three,crepieux_stm,dolcini_stm}.
It corresponds to a weak tunneling limit between the 1-D system to be probed and the electrodes. These can be either Fermi liquids, superconductors or non-Fermi liquids as well. For instance they are formed by multiwall nanotubes in Ref.\cite{gao_four}. In Ref.\cite{yacoby_fraction}, the electrode is itself a quantum wire under a controlled magnetic field which injects uni-directional electrons in another 1-D quantum wire. This experiment was devoted to show that an uni-directional electron splits into two solitons with partial charges. The phenomenon was predicted theoretically by the author\cite{ines_ann} and studied further by K. V Pham {\it et al}\cite{pham}, calling them "fractional charges" even they were not associated to any stable particles as in the FQHE. This experiment will be commented in a separate paper.

 A theory of all those multi-terminal setups need to take into account not only interactions in the arms, but especially the nature of boundary conditions and coupling at the junction. These play a key role and need to be treated in an appropriate way. In the Coulomb drag or crossed nanotubes setups, bosonisation has been used, and the coupling between two 1-D systems was shown to be essentially due to Coulomb interactions, while tunneling was claimed to be less relevant in the RG sense.
Inversely, in theoretical works dealing with electrodes weakly coupled to a one-dimensional system to be probed, such as STM or AFM, only tunneling  has been considered. This is usually justified by letting the distance between the wire and the electrode much higher than the typical Fermi wavelength. We aim in the present work to treat simultaneously the
 tunneling processes and the Coulomb interactions. In particular, tunneling will be shown to generate relevant processes. The interactions between the 1-D system and the electrode can induce similar processes, screening in addition the interactions inside the system.

 This raises questions about the weak tunneling limit, thought to ensure that one probes the intrinsic properties of the wire, not affected by the electrodes. These are expected to have a much less invasive effect than strongly coupled electrodes, which instead modify in a remarkable fashion both the thermodynamic and transport properties of the 1-D system\cite{ines_schulz,maslov_g,ines_ann}. Nevertheless, when taking into account second order tunneling processes as well as the possible Coulomb coupling, we show that even a weakly coupled electrode can have invasive effects which show up in the low-energy sector.

When the electrode is a sharp STM, we draw one crucial consequence of those invasive effects: when the wire or nanotube is grounded and the temperature is too low, it gets broken at the tunneling location. Thus the STM probes the end instead of the expected bulk DOS. The end DOS exponent is normally expected to be measured only at the end point of the system when it is weakly coupled to a reservoir. Its difference from the bulk DOS exponent has been used as an argument in favor of intrinsic interactions versus dynamical Coulomb blockade (DCB) \cite{ingold_nazarov}. DCB is a phenomena due to the fluctuations of the applied gate due the classical impedance of the circuit, and was shown to be equivalent to a local impurity problem in an infinite TLL \cite{ines_saleur}.  Thus invasive effects of the tip can hide one of the possible ways to discard it in favor of Coulomb interactions inside a finite TLL.

Many of the results we derive can be extended as well to the case where the electrode is an interacting system coupled by arbitrary spatial extension to the wire, thus are relevant as well for crossed or parallel 1-D systems as in Coulomb drag setups.

  Indeed, when the STM is sharp enough to be modeled by a semi-infinite non-interacting lead coupled by tunneling to a wire, it can be viewed as a special case of a junction between three TLL arms of different interaction strengths, made to vanish on the STM arm.
This interacting setup was treated theoretically in the limit of weak interaction strengths. Starting from a junction of three non-interacting arms described by a $3X3$ scattering matrix $S$\cite{sen_1}, the RG flow for $S$ was shown to have two sets of fixed points:
\begin{itemize} \item (a)-The completely disconnected junction.
\item (b)-One arm labeled $3$ weakly coupled to the perfectly connected two other arms (labeled by $2$ and $1$).
\end{itemize} The latter correspond to the STM setup when interactions vanish on arm $3$. By extending our analysis to the case arm $3$ is interacting, we show that the relevant second order tunneling process drives the "fixed point" (b) to (a). This makes it difficult to have (b) as a fixed point. Indeed, the RG approach to lowest order in interactions, given roughly by $1-g\simeq U/2\pi v_F$, is inappropriate at point (b). The bulk exponent in the TLL extending here continuously from terminal 1 to 2, is given by:
\begin{equation}
  \label{alphabulk}\alpha_{bulk}=\frac 12\left[\frac{1}g+g-2\right],
  \end{equation} and is of order $U^2$, thus cannot be treated by that approach. Point (a) was the fixed point of a Y junction  \cite{trauz_y} with arbitrary strength of interactions, but restricted to the same strength on each arm and to a fully symmetric scattering matrix: treating more general
conditions at the connection point within bosonisation has been a challenge. We believe that the framework developed here allows for progress in this direction.

In addition, many of the previous works\cite{sen_1,trauz_y} were restricted to the equilibrium regime where the transport is linear. Non-linear regime where acting on the three voltages can offer a more rich behavior. Crossed or parallel nanotubes were suggested  for instance to monitor voltage amplifiers \cite{egger_drag}. There have been works dealing with a sharp STM coupled by tunneling only to a TLL transport in an "out-of-equilibrium" situation, either in a ballistic wire\cite{crepieux_stm}, with one impurity \cite{lee_stm,eggert_dos,glazman_stm}. Imperfect contacts are studied in \cite{dolcini_stm,yacoby_theo}. A nice perturbative analysis of the current behavior in an STM or an AFM setup was given in Ref.(\cite{dolcini_stm}), taking into account finite-size effects due to the quasi-Andreev reflections \cite{ines_schulz,ines_epj,ines_ann,dolcini_05}. A sharp STM is shown to distinguish the latter from the standard reflections at imperfect contacts. Those effects are very important and interesting, and the formalism developed here allows their further investigations. But these effects will not be our main focus here.
Indeed we claim that one needs to go beyond the uniformly interacting TLL model considered in those works in order to include possible
 inhomogeneous screening effects due to the injector. Also we allow for spatial extension of all processes coming into play, including tunneling. This is important for the realistic setups. We offer as well a more general formalism for a systematic investigation of the current correlations. It applies also to a semi-infinite wire and can be extended easily to a finite wire with open boundaries coupled to many electrodes.

More importantly we draw attention to crucial features overlooked in those works. We have already mentioned the effective higher order tunneling processes, which requires caution when perturbation with respect to tunneling is performed as in these works.

Another crucial feature is related to a special case frequently used, letting the 1-D system grounded or semi-infinite as in \cite{eggert_dos,glazman_stm}. When the system is too long, one is left with two energy scales: the voltage of electron "injection" $V$ and the temperature $T$. These are usually expected to play symmetric roles as is the case in a two-terminal geometry. We show here that $V$ cannot ensure the two roles it is assigned usually: by letting $V\gg T$, one takes the zero-temperature limit, and by letting $V$ higher than a typical crossover energy we will denote by $\omega_{nq}^*$, one is still in the weak coupling regime. The subscript $nq$ indicates non-quadratic contribution to the Hamiltonian due to non-conserving momentum processes which can be already present, such as backscattering by imperfect contacts, or by the invasive effects we show. This means that the grounded wire is driven in the strong coupling regime when it is very long and at very low temperatures, even when $V>\omega_{nq}^*$. One consequence is that if the electrode is sharp and the wire has imperfect contacts, high $V$ cannot prevent the following fixed point: the 1-D system gets disconnected into two wires. Each of them is coupled at one end to the injector and the other to one reservoir. If the injector is non-interacting, both tunneling currents are controlled by the same end DOS exponent, Eq.(\ref{alphaend})!

This conclusion applies as well to the case the injector is itself an interacting wire. It applies also to three edge states in the FQHE regime we label by $1,2,3$. We can summarize our conclusion in this case as follows: a tunneling Hamiltonian for quasi particles between two edges, for instance $1$ and $2$, at the same voltage is relevant and cannot be treated perturbatively at zero temperature. The fact that the voltage diffrence between edges $1$ and $3$, $V_1-V_3$ is high does not permit the perturbation. $V_1-V_3$ is analogous to the voltage of our injector.
This shed light on the problems encountered by Kane and Fischer in Ref.\cite{kane_fisher_dilute}, motivated by an experimental work on shot-noise measurement\cite{heiblum_dilute}. They have noted that temperature and $V_1-V_3$ were not playing symmetric roles as they would expect. Nevertheless letting the zero temperature limit in that work was not justified, they had to keep $V_1-V_2$ finite for that.

Another important motivation for looking at multi-terminal geometries is to search for a signature of the statistics or other correlation effects in current noise correlations. The initial experiment of Hanbury-Brown and Twiss (HBT)
  held for photonic sources,\cite{hbt} has been proposed theoretically for fermions much
  later, and experiments have been achieved \cite{henny_hbt,oliver_hbt}.
It was shown that
cross correlations were always negative due to the Pauli
principle \cite{buttiker_prl,martin_landauer,blanter_buttiker}, or anti bunching of electrons.
Recently, there has been much interest in the search of deviations
from the negative sign, for instance through incoherent inelastic
scattering in integer edge states \cite{buttiker_texier}, hybrid
superconductor-normal systems
\cite{torres_martin,buttiker_hbt_supra}, and a quantum dot
 with ferromagnetic leads\cite{cottet_belzig}. In some of these works, the scattering approach could be
still used, which is no more possible in strongly correlated one-dimensional
systems. In
particular, elementary excitations in the FQHE at filling $\nu=1/(2n+1)$ have a fractional charge $\nu e$ which was measured through
shot noise in the poissonnian limit, more precisely the Fano factor given by $\nu e$ \cite{saminad,picciotto}. The two chiral edges can be described by a TLL with parameter $g=\nu$, analogous to a one-dimensional interacting wire. Nevertheless, the Fano factor of the wire cannot give access to $ge$ but only to $e$ \cite{ponomarenko_features}, unless one considers
high-frequency noise \cite{trauzettel_04}. This is due to the connection of a wire to charge reservoirs. It is related to its two-terminal DC conductance equal to to $e^2/h$ \cite{ines_schulz,maslov_g} instead of $\nu e^2/h$ in the FQHE.

This is one of the differences between the two systems, but there are also others when three-terminal geometries are considered such as the HBT setups. One
could wonder whether noise correlations could as well be used to reveal
deviations from Pauli statistics. The fractional statistics shown for Laughlin quasi-particles in the FQHE has not an analogue in the one-dimensional interacting wires. Though, interactions in the latter are expected to affect the cross-correlations in a non-trivial way.

On one hand, in order to find a signature of the fractional statistics in the FQHE, a first proposal of a three-terminal setup
was performed at simple fractional filling $\nu=1/(2n+1)$\cite{ines_prl}. This has been followed by a similar work
\cite{vish_hall}, has motivated an alternative geometry for
detecting telegraph noise
\cite{kane_telegraph}, and was extended for other fillings of the hierarchy states\cite{kim_hbt}. But contrary to wires or carbon nanotubes described by non-chiral TLL, the separation
of chiral edges allows a good control in order to favor fractional charge
tunneling instead of electrons, such that one could probe
fractional and not fermion statistics. However, no equivalent
set-up to that in \cite{ines_prl} could be achieved in non-chiral TLLs.

A different geometry proposed in \cite{crepieux_stm} was based precisely on a sharp STM coupled by weak tunneling to a TLL.
 Surprisingly, if the wire is infinite, homogeneous and grounded, without connexion to leads, and at zero temperature, cross correlations to lowest order in the electron
tunnel amplitude were positive. This was a paradox due to the fermion character of the injected electrons. In the presence of noninteracting leads the lowest order term vanishes \cite{crepieux_stm} leaving
open the question about the sign in a realistic measurement.
We will answer this question by showing that cross-correlations are negative. This is consistent with the Pauli statistics of the electrons which tunnel from the injector to the wire. Our result is valid more generally for any inhomogeneous and finite range interactions in the wire, and at different voltages in the three terminals.

The paper is organized as follows. Section \ref{summary} contains a summary of the main results, which does not respect necessarily the order of sections. Section \ref{twoterminal} deals with the two-terminal transport. It contains first a short review of the two-terminal transport in a TLL connected to reservoirs. Then we consider the more general inhomogeneous model, where interactions can be of arbitrary type and range. Subsection \ref{zeromodes} is concerned with the way the coupling to electrochemical potentials is treated, which will be of great use as well for the three-terminal geometry. In \ref{subcon} we expand formally the two-terminal differential conductance to all orders with respect to an extended backscattering potential. Then we give its scaling behavior to all orders for a local backscattering amplitude.
 In section \ref{formal}, we consider the case of an extended tunneling from the injector and write the general Hamiltonian where the three voltages of the injector and the two reservoirs are incorporated. We express the correlations of the current in a form which does not depend on the details of the Hamiltonian of the wire. Those formal expressions are suited for perturbation with respect to the tunneling amplitudes, done in section \ref{weaktunneling}. This section is central to the paper. It contains a discussion of the validity of perturbation with respect to relevant scattering processes in the wire, as well as the importance of limits between voltages and temperature. General results are derived for the auto- and cross-correlations in the weak tunneling limit, which are non-perturbative with respect to any relevant non-conserving processes in the wire. Even if initially absent from the initial wire, we show afterward that such processes can be generated by two invasive effects. The first effects are the virtual tunneling of electrons in two opposite paths, treated in section \ref{sectunnel}. The second ones are due to possible Coulomb interactions between the wire and the injector, treated in section \ref{invasive}, where screening effects are shown in addition. The generated processes are generally non-local in space and time, and an approximation is required to  write them as an effective Hamiltonian correction. In section \ref{sharp}, we apply those results to the case of a sharp electrode, relevant to a sharp STM for instance. Section \ref{scaling} contains a new behavior of the correlations of current in all orders of a local backscattering. It is relevant to section \ref{sharp} as well as to illustrate the results of section \ref{weaktunneling}.
 The last section, \ref{conclusion}, contains some concluding remarks and connection to other works.


 \section{Main results of the paper}
 \label{summary}

 Let's first present the Hamiltonian we adopt for the wire. We restrict ourselves to
spinless electrons, the extension to spinful electrons and many
channels as in carbon nanotubes is straightforward.
 The inhomogeneous TLL was intended not only to treat the attachment of leads with parameter $g_L$ to a wire with uniform parameter $g$ inside, but extended from the beginning to take into account arbitrary profile of $g(x)$ inside the wire, as well as finite-range interactions $U(x,y)$ and inhomogeneous scattering processes which do not conserve momentum \cite{ines_ann}. Thus it is suited to incorporate inhomogeneities due to either gates at non-uniform distance, fluctuations in the confining potential, and, most importantly for our present purpose, the invasive effects of an injector discussed in sections \ref{sectunnel} and \ref{invasive}. The latter occur through both inhomogeneous screening and non-local backscattering processes generated by second order tunneling and Coulomb interactions between the wire and the injector, shown in \ref{sectunnel} and \ref{invasive}. We will as well propose a possible and analogous microscopic treatment of a close gate to include its screening effects.
 Thus throughout the paper, unless specified, the wire has inhomogeneous Coulomb interactions of arbitrary form and range described by a quadratic bosonized Hamiltonian $\mathcal{H}_{q}$, Eq.(\ref{H0}), where $g(x)$ and $U(x,y)$ implicitly include the screening effects shown in section \ref{invasive}. It has as well non-conserving momentum processes encoded in a non-quadratic Hamiltonian $\mathcal{H}_{nq}$, which implicitly contain the non-local and inhomogeneous backscattering terms generated by both tunneling to order $\Gamma_+\Gamma_-$, Eq.(\ref{HnqT}) and those due to Coulomb interactions, Eqs.(\ref{HnqCoul}). The total Hamiltonian of the wire $\mathcal{H}_W$ is given by Eq.(\ref{Hdecomposed}).

  Reservoirs can be simulated by boundary conditions on operators simulating "chemical potentials" at the ends of a finite wire \cite{ines_epj}, see Eq.(\ref{boundaryconditions}). These are similar to the so called radiative boundary conditions, which are imposed on averages intsead of operators \cite{egger,egger_err} and obtained in the infinite wire limit.  An elegant proof for generalizing the DC conductance result $e^2/h$ was derived in Ref.(\cite{ines_epj}). It requires simply the conservation of the total charges for the right and left-going electrons, whatever the type of interactions are inside the system. Here we reproduce it for the specific case of a quadratic Hamiltonian, Eq.(\ref{H0}) (see section \ref{zeromodes}). Indeed the boundary conditions on operators, Eq.(\ref{boundaryconditions}), were shown to be equivalent to attach the TLL to noninteracting leads with constant potentials, Eq.(\ref{profile}). If the leads themselves were as well TLL liquids with a different parameter $g_L$, the DC conductance was shown to be given by the latter, $g_Le^2/h$, whatever the profile of interactions inside the wire are \cite{ines_bis}. Our formalism applies as well to a semi-infinite wire, letting in particular $g_L=0$.
 \begin{enumerate}

 \item  We implement the boundary conditions Eq.(\ref{boundaryconditions}) to derive the zero modes for any range and profile of interactions in \ref{zeromodes}. This gives a different analysis from that in Refs.\cite{ponomarenko_features,dolcini_stm} for the short-range interacting case. It sheds light on the appearance of the sum of the voltages at the reservoirs $V_1+V_2$ without recourse to the gate.
 In particular, we show that with three voltages, there are two types of energy scales which play a crucial role, besides the length of the wire and temperature (see Eqs.(\ref{omega12},\ref{omega(x)})):
\begin{eqnarray}\label{thetwo}
\omega(x)&=&\frac{e}{2\hbar}[V_1+V_2-V(x)]\nonumber\\
 \omega_{12}&=&\frac{e g_L}{2\hbar}(V_1-V_2),
 \end{eqnarray}  where $g_L$ is the interaction parameter of the leads and $\omega_{12}$ is independent on $x$.
 $\omega(x)$ depends on each tunneling point $x$. We will not introduce a frequency associated to temperature $T$ to avoid confusion, thus whenever $T$ is compared to other frequencies, it is $k_BT/\hbar$ which is meant. The potential $V(x)$ is determined in a microscopic way (Eq.(\ref{V(x)})), taking into account some interactions inside the wire as well as with the gate and the injector. No self-consistent determination has to be done here.

\item
  We give the energy dependence the two-terminal DC differential conductance for spatial extended impurities in a finite wire.  This could simulate for instance imperfections at the contacts, as well as the backscattering generated by extended tunneling or interactions with the injector. We will offer formal expansions which give the qualitative dependence of current correlations in the weak backscattering regime (WBS) in some important limits, see Eqs.(\ref{coherent},\ref{incoherent}). We give as well general scaling laws in Eq.(\ref{scaling12}) for both WBS and strong backscattering regime (SBS) regime. They will serve as well in a future work commenting the experiment of Ref.\cite{yacoby_fraction}.

 \item
 The current correlations are expressed formally and exactly in the presence of an injector at an electrochemical potential $\mu_3=eV_3$, see Eqs.(\ref{I12},\ref{S12}).  The currents operators are taken at the contacts to the leads. Tunneling to the wire is spatially extended, and asymmetric amplitudes $\Gamma_{\pm}(x,\vec{r})$ for right and left-going electrons are adopted. Here $x$ is in the wire and $\vec{r}$ in the injector. This allows more general situations as in \cite{yacoby_fraction,dolcini_stm}. The amplitudes $\Gamma_{\pm}$ are as well multiplied by two asymmetric time-oscillating factors $e^{i\omega_r(x)t}$ with $r=+$ ($r=-$) for right (resp. left-going) electrons, where (see Eqs.(\ref{thetwo},\ref{Vr},\ref{V(x)}))
 \begin{equation}
 \omega_r(x)=\omega(x)-r\omega_{12}.
 \end{equation}

 Then we express the current correlators to lowest order in tunneling (section \ref{weaktunneling}) but still non-perturbatively with respect to $\mathcal{H}_{nq}$. We will see that this is justified as the generated relevant process in $\Gamma_+\Gamma_-^{\dagger}$, Eq.(\ref{HnqT}), is implicitly included in $\mathcal{H}_{nq}$. The zero-order contribution to current correlators with respect to tunneling is given by their exact two-terminal values in the presence of $\mathcal{H}_W$. In particular, when $\mathcal{H}_{nq}=\mathcal{H}_{B}$ describes backscattering on an arbitrarily extended potential $\lambda(x)$, the two-terminal differential conductance expansions of \ref{subcon} can be used.

 \item

  A crucial result we reveal is the following. Consider the currently used situation of an infinite grounded TLL coupled by weak tunneling to a tip at a voltage $V$, which plays the role of $\omega(x)$ when $\omega_{12}=0$. Then one is left with only $V$ and $T$ which are expected to intervene through a scaling law. In particular the correlations of the currents depend only on $max(V,T)$ if either $V\gg T$ or $\ll T$.

We will show that such an expectation is not appropriate. When the wire is grounded, i. e. letting:
\begin{equation}\label{grounded}
\omega_{12}=0\Rightarrow\omega_+(x)=\omega_-(x)=\omega(x),
\end{equation}  we show that $\omega(x)$ cannot play the roles one usually attributes to it:
\begin{itemize}
\item It does not provide a cutoff for the RG flow of $\mathcal{H}_{nq}$.
\item Letting $\omega(x)\gg T$ does not allow to take the zero-temperature limit in a perturbative expansion of the correlations of currents with respect to $\mathcal{H}_{nq}$.
\end{itemize}

This allows to draw two related conclusions:
\begin{itemize}
 \item That the validity of a perturbative expansion with respect to $\mathcal{H}_{nq}$ is the same as in a two-terminal geometry, i. e. that:
 \begin{equation}\label{omega*}
 \omega^*=max(T,\omega_{12},\omega_L)>\omega_{nq}^*.
 \end{equation}
Here $\omega_{nq}^*$ is the typical crossover frequency to the strong scattering regime associated to $\mathcal{H}_{nq}$, and
\begin{equation}\label{omegaL}
   \omega_L=\frac v L,
   \end{equation}
  $v$ being the plasmon velocity.
 \item That the precise out-of-equilibrium condition, in the sense that the zero-temperature limit can be undertaken in such perturbative expansion, is given by:
\begin{equation}\label{out0}
\omega_{12},\omega(x)\gg k_BT/\hbar. \end{equation}
 \end{itemize} Thus if $\omega^*<\omega_{nq}^*$, letting $\omega(x)>\omega_{nq}^*$ does not prevent the wire from being driven to the strong coupling regime corresponding to $\mathcal{H}_{nq}$. For instance if one has ohmic contacts where weak backscattering holds, they are driven to tunneling contacts. But even if one starts from a pure wire, with a Hamiltonian given by $\mathcal{H}_q$ only, tunneling as well as possible interactions between the injector and the wire generates an effective $\mathcal{H}_{nq}$, see sections \ref{sectunnel} and \ref{invasive}. Besides, the voltage of the injector cannot prevent the wire from going to the associated strong coupling regime.

\item More precisely, we show that the virtual process of tunneling at position $x_1$ followed by the tunneling in the opposite sense of at position $x_2$ generates effective non-local backscattering processes from $x_1$ to $x_2$, with an amplitude proportional to $\Gamma_+\Gamma_-^{\dagger}$. Those add to the Hamiltonian $\mathcal{H}_{nq}$, see Eq.(\ref{HnqT}). Notice that this effect is absent for uni-directional injection. It can already be guessed from the non-interacting
     case because of the unitarity of the $3X3$ scattering matrix. In particular, even if the wire is described by the quadratic Hamiltonian $\mathcal{H}_{q}$ only, the usual lowest-order approach with respect to tunneling is not sufficient, as the next order term is more relevant. Thus the generated processes have to be included in $\mathcal{H}_{nq}$, as we do throughout the paper.

     This result applies as well if the injector is itself an interacting wire, and has therefore to be taken into account for Coulomb drag or crossed wires or nanotubes. As mentioned in the introduction, it questions the fixed point in \cite{sen_1} where an interacting wire is weakly coupled to the bulk of another one. (section \ref{sectunnel} and \ref{sharp})
 \item  We discuss another possible invasive source, even though less systematic: Coulomb interactions between the injector and the wire allowed as well to have arbitrary spatial extension. This is the subject of section \ref{invasive}. While most of the formalism and results apply to the case the injector is itself a one-dimensional interacting wire, this section cannot be rigorously extended to include this situation, unless some conditions to be determined are ensured. This is because we adopt the hypothesis that the density fluctuations in the injector are plasmonic modes commuting with the vertex operators. Thus they don't include the higher harmonic components one has in a 1-D injector, and retained for instance in Coulomb drag or crossed 1-D systems works. Nevertheless, these works have ignored the contribution we treat, which might be relevant too.

 Here we show that Coulomb interactions between the injector and the wire have mainly three possible consequences whose relative importance depend on the precise geometry, spatial extension and distance of the injector to the wire:
\begin{itemize}\item Backscattering terms due to the screened electric potential when it varies on the scale of few wavelengths.
 \item Non-local backscattering processes similar to those generated by tunneling and affecting $\mathcal{H}_{nq}$, see Eq.(\ref{HnqCoul}).
\item Inhomogeneous screening of the interactions inside the wire, which affects the functions $g(x)$ and $U(x,y)$ in $\mathcal{H}_q$, Eq.(\ref{H0}).
\end{itemize}  Screening of the interactions can affect the properties of the wire in a more considerable way for a large spatial extension of the injector, whereas the generated backscattering type processes can play their most important role for sharp injectors, especially if made close enough to the wire.

 \item We specify those results to a sharp injector.  As it induces local backscattering type terms due to the invasive effects, the strong coupling regime corresponds to a wire disconnected at the tunneling location. Besides, the sharp voltage of the injector cannot prevent from going into this regime. In such limit, if the sharp injector is an STM, it probes the end DOS instead of the bulk one! Those are given respectively by Eq.(\ref{alphaend}) and (\ref{alphabulk}) if the wire is a homogeneous TLL with parameter $g$. As this difference between the end and bulk DOS exponent is invoked to discard the DCB, such result needs to be taken into consideration.
 Notice that if the wire had initially almost good contacts, the wire gets disconnected both at its ends and at the tunneling location in the common associated SBS. (section \ref{sharp})

   \item

     We analyze current auto- and cross-correlations. The latter can yield a signature of charge statistics or correlation effects in HBT geometries. In order to achieve such purpose, one needs the zero-temperature limit to get rid of thermal fluctuations. At the same time, it is frequent to let two terminals at the same voltage, as in \cite{crepieux_stm}. We show here that this is not adequate, and that we need to ensure the criteria in Eq.(\ref{out0}) for that. We express formally and non-perturbatively the current correlations for any extended tunneling amplitudes in section \ref{formal} and for weak extended tunneling amplitudes in \ref{weaktunneling}. Then we show that given any Hamiltonian with relevant scattering processes in $\mathcal{H}_{nq}$, the cross-correlations are simply dominated by:
$$S_{12}\simeq -s_{nq,nq}$$ where $s_{nq,nq}$ is the exact out-of-equilibrium noise in the two-terminal geometry.

When the Hamiltonian density in $\mathcal{H}_{nq}$ is local in space, $s_{nq,nq}$ is positive, thus $S_{12}<0$. In a pure wire coupled locally to a sharp STM, where initially $\mathcal{H}_{nq}=0$, the generated backscattering due to second-order tunneling events dominate and yields a negative sign. Coulomb interactions with the injector can as well contribute to $s_{nq,nq}$.
In particular, we solve a paradox in \cite{crepieux_stm} related to the cross-correlations. Though they should be negative as electrons obeying Fermi statistics are injected, they were shown to be positive instead. This result was obtained in an infinite grounded homogeneous TLL coupled by local tunneling to a sharp tip.

 We make in addition a general and simple observation, which can be of great use: one advantage of considering noise in a three-terminal geometry keeping two finite voltages is to define in two possible ways "excess" noise. This is done by subtracting the noise when one of the two voltages vanish, for instance when $\omega_{12}=0$ as in Eq.(\ref{excess}). This allows to get rid of any undesirable background noise, similarly to the case when considering two-terminal finite frequency noise.
 \item
In section \ref{scaling}, we illustrate explicitly the previous results in the limit of an infinite wire and for both local backscattering potential and tunneling. This is in particular relevant for a sharp STM in view of the invasive effects.
We give generalized and new scaling laws with respect to the three energy scales, $\omega(x)$, $\omega_{12}$ and $T$ in Eq.(\ref{scaling12}). In particular, when taking the limit of a grounded wire, i. e. at $\omega_{12}=0$, we show clear deviations from the expected scaling laws with respect to the voltage of the injector and the temperature. Currents and noise are expanded in a formal way to all orders in backscattering: temperature and voltages don't play anymore symmetric roles as in the two-terminal case, Eq.(\ref{scaling12}). This might be relevant to measured deviations in Ref.\cite{bockrath_99}. Our analysis applies more generally to three interacting arms. It could as well be applied to three edges in the FQHE, and sheds light on peculiarities encountered in Ref.\cite{kane_fisher_dilute}.

\end{enumerate}

 \section{The two-terminal geometry}
 \label{twoterminal}
 The first subsection contains a review of a TLL with uniform short-range interactions parametrized by $g$ connected to leads, and with one weak or strong impurity. In subsection \ref{general}, we consider arbitrary form and range of interactions encoded in the quadratic part $\mathcal{H}_{q}$, and any scattering processes which do not conserve momentum encoded in the non-quadratic part $\mathcal{H}_{nq}$, including for instance backscattering at the contacts. Then we derive and interpret some features of the transport for the wire connected to two reservoirs at electrochemical potentials $\mu_1$ and $\mu_2$. These are implemented by two equivalent alternatives recalled in subsection \ref{zeromodes}. In particular, the perfect conductance $e^2/h$ result was derived more generally when charges for right and left movers are conserved, without specifying the type of interactions inside the wire \cite{ines_epj}. This proof will be applied here for the quadratic Hamiltonian $\mathcal{H}_{nq}$. The main aim of subsection \ref{zeromodes} is to implement the zero modes in the out-of-equilibrium situation, using the concepts of Ref.\cite{ines_epj} to get a universal and transparent way. In \ref{subcon}, seeking definiteness, we discuss the transport in case where $\mathcal {H}_{nq}=\mathcal{H}_{B}$ describes backscattering on an extended potential profile $\lambda(x)$. Then we expand the two-terminal differential DC conductance $G_{12}$ and noise formally in powers of $\lambda(x)$, the noise has as well similar behavior. This will be useful for the expansion of the currents and noise in the presence of the injector. In subsection \ref{scaling} we give the formal scaling behavior to all orders of the DC conductance for a local backscattering in an infinite wire.

    \subsection{Brief review of the TLL model}
    \label{reviewTLL}
 A pure TLL has a DOS depending in power law on energy, the exponent being given by Eq.(\ref{alphabulk}). It has been established that it is more sensitive to disorder compared to higher dimensions. For Gaussian extended disorder Giamarchi and Schulz showed that impurities induce localization at $g<2/3$, and induce renormalization of $g$ itself\cite{giamarchi_loc}. For one or two local impurities, Kane and Fischer\cite{kane_fisher} showed that at low enough energy the conductance is suppressed again. A weak impurity strength given by $\lambda$ gets renormalized as:

 \begin{equation}\label{lambda*}
 \lambda^*(\omega^*)\simeq \lambda\left(\frac{\omega^*}{\Lambda}\right)^{g-1} ,
 \end{equation}
  obtained by cutting the RG flow at the IR frequency:
 \begin{equation} \label{omegatwo*}
\omega^*=max(k_BT/\hbar,|V_1-V_2|,\omega_L),
 \end{equation}
  where $\Lambda$ is a typical energy cutoff, and  $\omega_L$ given by Eq.(\ref{omegaL}).
   The finite length of the wire was not implicitly taken into account in those previous works, but enter through this scaling argument. The perturbative correction to the differential DC two-terminal conductance reads:
  \begin{equation}\label{deltaG12}
  \delta G_{12}\simeq |\lambda^*(\omega^*)|^2.
  \end{equation}

  For repulsive interactions, $g<1$, $\lambda^*$ diverges at $\omega^*$ smaller than :
  \begin{equation}\label{omegaB*}
\omega_B^* \simeq  \Lambda |\lambda|^{2/(1-g)}.
 \end{equation}
  This is the crossover energy from weak backscattering (WBS) to the strong backscattering regime (SBS). For $\omega^*<\omega_B^*$, the conductance vanishes in the limit of vanishing $\omega^*$. In this limit the wire gets disconnected into two semi-infinite wires at the impurity location. One could  perform perturbative computation with respect to a weak tunneling amplitude $t_{12}$ from one end-point to the other, the renormalized tunneling amplitude reads now:

 \begin{equation}\label{t12*}
t_{12}^*(\omega^*)\simeq t_{12} \left(\frac{\omega^{*}}{\Lambda}\right)^{(1/g-1)},
\end{equation}
and the differential conductance
\begin{equation}\label{G12tunnel}
G_{12}\simeq |t_{12}^*(\omega^*)|^2
\end{equation} vanishes for $\omega^*\rightarrow 0$ as mentioned.
In those both limits, the shot-noise at zero-temperature has the same behavior as in Eqs.(\ref{deltaG12},\ref{G12tunnel}), multiplied  respectively by the charge $ge$ or $e$ and the bias $V_1-V_2$.

  For an infinite wire, it was possible to perform two series expansions for the current cumulants: one valid below $\omega_B^*$, i. e. in the SBS, and the other above, in the WBS regime \cite{fendley}.

    It has been well established nowadays that the features of quantum wires or single wall carbon nanotubes get modified by connection to reservoirs.
The DC conductance $G_{12}$ of a pure wire was thought initially to be determined by $g$ and thus to offer a "simple" way to measure $g$ \cite{apel_rice,kane_fisher}. Taking into account the connection to charge reservoirs yields instead $G_{12}=e^2/h$ \cite{ines_schulz,maslov_g}.

    Nevertheless, $g$ still enters in a non-trivial way in the AC conductance $G_{12}(\omega)$\cite{ines_schulz,ines_epj,blanter} or in various transport features in the presence of impurities\cite{ines_nato,maslov_disorder,furusaki_fil_fini,dolcini_05,trauzettel_04,ines_ff_noise}. In particular, the typical hallmark of the TLL model is not washed out by the presence of the reservoirs: power law energy dependence of the DOS or the backscattering current arises, with an exponent determined by $g$. However, this could be as well a signature of dynamical Coulomb blockade (DCB), a phenomenon occurring in a coherent conductor in series with an ohmic circuit whose voltage fluctuations reduce the current through the conductor as well in a power law fashion.\cite{ingold_nazarov} This raises doubts on a strong evidence for the TLL theory, especially that the two problems have been shown to be equivalent.\cite{ines_saleur} Nevertheless, the mapping holds when the finite size of the wire is not taken into account: this allows for the finite size effects to offer ways to distinguish the role of intrinsic interactions from an external resistance.
  A first way is through typical oscillations with a period inversely proportional to $\omega_L$ (Eq.(\ref{omegaL})) due to a phenomenon caused by variations of the interaction parameters and called multiple quasi-Andreev reflection at ohmic contacts \cite{ines_schulz,note_oreg}. Oscillations show up in the AC linear\cite{ines_epj} and AC differential conductance,\cite{ines_ff_noise} in the non-linear current \cite{dolcini_05} as well as in finite-frequency noise in the presence of an impurity\cite{dolcini_05,trauzettel_04,ines_ff_noise}. They are superimposed on the power law behavior. This gets however more complicated due to the breaking of translational invariance, which offers precisely a second way to discard the DCB phenomena, even though this argument still requires a rigorous settlement. For instance, in the exponent for the DOS at the bulk given by Eq.(\ref{alphabulk}),
  $g$ has to be replaced by an effective "contact" parameter $g_c$ for the DOS at one ohmic contact. For $g_L=1$, it is given by \cite{ines_schulz,ines_nato}:
  \begin{equation}\label{gc0}
  g_c=\frac{2g}{1+g}
  \end{equation}  Indeed $g_c$ is the transmission coefficient of a soliton charge incident from a reservoir at the contact. Similarily, a weak impurity reduction to the conductance at an ohmic contact is given by Eqs.(\ref{lambda*},\ref{deltaG12}) where one has instead the exponent $g_c-1$. In the SBS regime, tunneling between the wire and the noninteracting lead is rather controlled by the dual parameter:
  \begin{equation}\label{alphaend}
  \alpha_{end}=\frac1{g_c}-1=\frac 12\left[\frac{1}g-1\right].
  \end{equation} Thus one recovers the end DOS exponent in the weak tunneling to a reservoir starting from open boundaries.\cite{fabrizio_open}

  Letting $g_L\neq 1$, one could get as well the tunneling exponent between two TLL with different parameters $g$ and $g_L$. For perfect contacts, the transmission coefficient of a charge soliton is, instead of Eq.(\ref{gc0}), given by \cite{ines_schulz}:

 \begin{equation}
 g_c=1-\gamma=\frac{2g}{(g_L+g)},\label{gc}
 \end{equation}
and
 \begin{equation}\label{gamma}
 \gamma=\frac{g_L-g}{g_L+g}
 \end{equation}
 is the coefficient of what we called "quasi-Andreev" reflection \cite{ines_schulz,note_oreg}.
 An impurity at the contact is now controlled by $g_Lg_c-1$. Thus the tunneling regime between the two TLL is controlled by the dual exponent $$\alpha=\frac 1{g_Lg_c}-1=\frac 1 2\left[\frac 1 g+\frac 1 {g_L}-2\right].$$ In particular, one recovers Eq.(\ref{alphaend}) for $g_L=1$, and the exponent in Eq.(\ref{t12*}) for $g=g_L$.

   As the finite size has been explicitly taken into account, $\omega_L=v/L$ does not only offer a possible cut off for the RG flow in Eq.(\ref{lambda*}). Recall that the quasi-Andreev reflection phenomenon was developed for chiral right and left propagating plasmonic density operators \cite{ines_schulz,ines_ann,pham} defined by:
 \begin{equation}\label{Phichiral}
 \tilde{\rho}_r=\frac{rg(x)+1}2 \rho_+-\frac{rg(x)-1}2 \rho_-,
 \end{equation}
   where $r=\pm$ and $\rho_{\pm}$ are the densities for right and left going electrons.
   Thus $\tilde{\rho}_\pm(x\mp vt)$ propagate at the velocity $\pm v$ as long as $g(x)$ is uniform. When a bare electron is injected inside a uniform TLL, it splits immediately into two opposite chiral charges $(1+g)/2$ and $(1-g)/2$\cite{ines_ann,pham} propagating at the plasmon velocity $v$. Nevertheless, those two soliton charges get reflected back and forth once they reach the contacts.

    If instead of a local impurity one adds an extended Gaussian disorder, perturbative computation for the finite wire allowed to recover the localization length \cite{ines_nato,ines_ann,ines_prb_long} found by Giamarchi and Schulz using a scaling argument \cite{giamarchi_loc}, thus the localization occurs at $g<3/2$. For a fixed extended disorder, the transition was shown to occur even earlier, $g<2$ \cite{ines_ann}. But contrray to the case of one or few local impurities, the interactions are renormalized by an extended backscattering potential.

   \subsection{A general bosonized Hamiltonian with coupling to reservoirs and DC conductance}
   \label{general}
 Here we consider a general inhomogeneous one-dimensional system in the low-energy sector, not necessarily described by the TLL model discussed above. Thus the the spectrum can be still linearized around the two Fermi points $\pm k_F$, thus to introduce the density for right and left-going electrons $\rho_{\pm}$. The long-wavelength part of the density can be expressed through:
 \begin{eqnarray}  \label{rho}
\rho(x)&=&\rho_+(x)+\rho_-(x)
\end{eqnarray}

 One can also introduce two bosonic fields $\Phi$ and $\Theta$ obeying  $[\Theta(x),\Phi(y)]= i sgn(x-y)$, such that:
\begin{eqnarray}  \label{rhobis}
\rho(x)&=&-\frac 1{\pi}\partial_x\Phi(x)
\end{eqnarray}

The Fermi fields for the right and left-going electrons labeled by $r=\pm$ are given by:
\begin{eqnarray}\label{Psir}
\Psi_r(x)&=&\frac{1}{\pi a}e^{irk_F x-i\Phi_r(x)}\nonumber\\
\Phi_r(x)&=&r\Phi(x)-\Theta(x)\nonumber\\
\rho_r&=&-\frac 1 \pi \partial_x \Phi_r,
\end{eqnarray}
where $a$ is a short-distance cutoff roughly given by $v_F/\Lambda$, $\Lambda$ being a typical energy bandwidth. Notice that $\Phi_r$ is conjugate to $\rho_r$. The total density operator $\rho_{tot}$ contains, besides $\rho$, $2mk_F$ components\cite{haldane_bosons}. We adopt here a
full expression which ensures particle conservation, such that
an uniform potential has no effect:
\cite{ines_resonance,sablikov_cdw}:
\begin{equation}
\label{rhotot}\rho_{tot}(x)=\rho(x)-\partial_x\left(\sum_{m=-\infty,m\neq
0 }^{\infty}\frac{1}{2i\pi m}e^{2im(\Phi(x)-k_Fx)}\right).\end{equation}

As any interactions imply the fields $\Psi_r$ or the total density $\rho_{tot}$, the total Hamiltonian of the 1-D system $\mathcal{H}_W$ can be viewed equivalently as a functional of one of the pairs: $(\rho_+,\rho_-)$ or $(\rho,j)$ or $(\Phi,\Theta)$.

A first way to implement the connection to reservoirs is to impose boundary conditions on fields suggesting chemical potentials for right and left-going electrons. They are defined by \cite{ines_epj}:

\begin{equation}\label{mupm}
\mu_{\pm}(x)=\frac{\delta {\mathcal{H}}}{\delta \rho_{\pm}(x)}.
\end{equation}
We reproduce as well the universal expression of the current given by:
\begin{eqnarray}\label{juniversal}
j(x)&=&\frac e h [\mu_+(x)-\mu_-(x)].
\end{eqnarray}
A local "average" chemical potential field was as well introduced:
\begin{eqnarray}\label{mu}
\mu(x)&=& \frac{\delta \mathcal{H}}{\delta \rho(x)}=\frac{1}2 [\mu_+(x)+\mu_-(x)],
\end{eqnarray}
and plays a symmetric role to the current in Eq.(\ref{juniversal}).

For a finite wire with ohmic contacts at $\pm L/2$ connected to reservoirs at electrochemical potential $\mu_1$ and $\mu_2$, one imposes:
\begin{eqnarray}\label{boundaryconditions}
\mu_+\left(-\frac{L}2 ,t\right)&=&\mu_1\nonumber\\
\mu_-\left(\frac{L}2 ,t\right)&=&\mu_2.
\end{eqnarray}
This implementation allows as well to treat time-varying $\mu_l(t)$ \cite{ines_epj}.

The second way was provided by attaching noninteracting leads. It is equivalent to those conditions provided one adds to the Hamiltonian of the wire+leads a coupling to a potential $V(x)$:
\begin{equation}\label{coupling}
\mathcal{H}_V=e\int V(x)\rho(x).
\end{equation}
An important condition to get the previous boundary conditions is to let $V(x)$ constant on the leads \cite{ines_bis,ines_ann,ines_epj,ponomarenko_features}:
\begin{equation}
V(x) = \left\{
\begin{array}{ll}
V_1=\mu_1/e & \mbox{for } x < -\frac{L}{2} \\
& \\
V_2=\mu_2/e& \mbox{for }  x > + \frac{L}{2} \\
\end{array}
\right. \label{profile}
\end{equation}Thus the applied voltage is
\begin{equation}
V_{12} \,\doteq \,(\mu_1-\mu_2) \, / \, e  \; . \label{def-voltage}
\end{equation}
 The profile of $V(x)$ inside the wire is the same for both ways of implementing the reservoirs. It is discussed in \ref{chemicalgate} in the two-terminal geometry taking into account the presence of a gate, as well as in \ref{chemicalinj} when the effect of the injector is included.

  We will pursue in the present and next subsection with the first way, thus conditions in Eq.(\ref{boundaryconditions}), as they are conceptually more transparent and clarifies in particular the voltage combination which will enter in the tunneling Hamiltonian later. For explicating current correlators, the model with leads will be more convenient instead.

Let us be more specific now, and separate the total Hamiltonian into a quadratic and non-quadratic part.
The quadratic part can be in general written as (see Eq.(\ref{rhobis})):
\begin{eqnarray} \label{H0}
{\mathcal{H}}_q &=&\frac{h}{2}  \int
 \frac{dx} { v(x)g(x)}\left\{  [\partial_x\Theta(x)]^2  + v(x)^2
[\partial_x\Phi(x)]^2\right] \nonumber\\+&&\int \frac{dx dy}{\pi ^2} U(x,y)\partial_x\Phi(x)\partial\Phi(y).
\end{eqnarray}
where $v$ and $g$ are
interaction parameters in which the interactions $U$ can be absorbed when they become short-range. In this case the parameters $v(x),g(x)$ become similar to the plasmon velocity and usual interaction parameter $g(x)$. The product $vg$, called the charge stiffness, is quite often given by : $vg=v_F$ when Galilean invariance has to be ensured. This is not however systematically the case; for instance it can be renormalized by umklapp process or electron-phonon interactions in the domain where they are irrelevant.
Besides $\mathcal{H}_q$ in Eq.(\ref{H0}), one can have other scattering processes which do not conserve momentum and which do not reduce to a quadratic form. They are described generically by a non-quadratic Hamiltonian $\mathcal{H}_{nq}$, which has often an exponential dependence on the bosonic fields. Spatial inhomogeneous processes enter as well in $\mathcal{H}_{nq}$. Possible terms due to these processes can be linear in $\rho$ and are incorporated in $V(x)$, Eq.(\ref{coupling}) as we will illustrate later. Thus the total Hamiltonian of the wire can be expressed as:
   \begin{equation}\label{Hdecomposed}
   \mathcal{H}_W=\mathcal{H}_q+\mathcal{H}_{nq}+\mathcal{H}_V.
    \end{equation}

    We aim to show now how both coupling to reservoirs at $\mu_1,\mu_2$ and to the internal potential $V(x)$ are incorporated by translation of the fields $\Phi$ and $\Theta$, thus by adding zero modes contributions. This is done in the presence of $\mathcal{H}_q$ only, afterward $\mathcal{H}_{nq}$ will incorporate these translations. As the boundary conditions in Eq.(\ref{boundaryconditions}) make this issue more transparent, we assume so far the integrals are over the finite wire $-L/2,L/2$. When leads are attached, the integrals in Eq.(\ref{H0}) run rather over the infinite system wire and leads, and $v,g$ are taken to be constant equal to $v_L,g_L$ on the leads, as well as $V(x)$ in Eq.(\ref{profile}).

    \subsubsection{Quadratic Hamiltonian and zero modes}
    Here we pick in Eq.(\ref{Hdecomposed}) only the Hamiltonian:
    \begin{equation}\label{H0bis}\mathcal{H}_0=\mathcal{H}_q+\mathcal{H}_V,
    \end{equation}
   where $\mathcal{H}_q$ is explicated in Eq.(\ref{H0}).
     We apply some other features of Ref.(\cite{ines_epj}) to this case. As the Hamiltonian is now quadratic, equations of motion are sufficient for our purpose. Recalling that the right and left bosonic fields
 $\Phi_{\pm}$ in Eq.(\ref{Psir}) are conjugate to the right and left-going densities $\rho_{\pm}$, and in view of the definition of $\mu_{\pm}$, Eq.(\ref{mupm}) one has :
\begin{equation}\label{motionmur}
\partial_t\Phi_r=-\frac 1 {\hbar}\mu_r.
\end{equation}
Thus the fields
\begin{eqnarray}\label{phitheta}
\Phi &=&\Phi_++\Phi_-\nonumber\\
\Theta &=&\Phi_+-\Phi_-
\end{eqnarray}
obey, using Eqs.(\ref{juniversal},\ref{mu}):
\begin{eqnarray}\label{equationmotion}
\partial_t\Phi&=& \pi j\nonumber\\
\partial_t\Theta &=&\pi \mu.
\end{eqnarray}
Thus the zero-mode parts depending on time in both $\Phi$ and $\Theta$ can be found by finding the stationary values of $j$ and $\mu$, or those of $\mu_{\pm}(x)$. For this, let's differentiate Eq.(\ref{motionmur}) with respect to $x$:
\begin{equation}
h\partial_t\rho_r=\partial_x \mu_r.
\end{equation}
 In particular, in the stationary limit:
\begin{equation}
\partial_x\mu_{r,0}(x)=0.
\end{equation}
This allows, in view of Eq.(\ref{boundaryconditions}), to obtain:
\begin{eqnarray}
\mu_{+,0}(x)&=&\mu_1\nonumber\\
\mu_{-,0}(x)&=&\mu_2.
\end{eqnarray}

 Thus the stationary value of the fields $\mu$ and $j$ in Eqs.(\ref{juniversal},\ref{mu}) are given by:

\begin{eqnarray}\label{stationarylimit}
\mu(x)&=&\mu_0=\frac 1 2 [\mu_1+\mu_2]\nonumber\\
j(x)&=&I_0=\frac e h [\mu_1-\mu_2].
\end{eqnarray}

Now Eqs.(\ref{equationmotion},\ref{stationarylimit}) allow to get the zero-mode parts proportional to time:
\begin{eqnarray}
\Phi&\rightarrow& {\pi} I_0t\nonumber\\
\Theta &\rightarrow&  \pi\mu_0 t.
\end{eqnarray}

Besides, the relations in Eq.(\ref{stationarylimit}) shed light on two issues.
 Firstly, as $\mu_1=eV_1$ and $\mu_2=eV_2$, we can deduce that the DC conductance is $e^2/h$ for the Hamiltonian in Eq.(\ref{H0}). This result could as well have been obtained through connection to noninteracting leads, and using Eqs.(\ref{coupling},\ref{profile}). More generally, without specifying the Hamiltonian, it was shown that if the total right and left charges are conserved, the DC conductance is still equal to $e^2/h$ \cite{ines_epj,ines_bis}. In this general case the uniformity of $\mu_{\pm}(x)$ is not ensured inside the wire, but one has $\mu_{+,0}(\pm L/2)=\mu_{1,2}$.

 Letting the leads interacting for generality, with TLL parameter $g_L$, one has also:
\begin{equation}\label{G12q}
G_{12}^q=\frac {e^2} h g_L.
\end{equation}
Secondly, equation (\ref{stationarylimit}) shows that it is not only the voltage difference which matters, but the sum of the two voltages: it is the average of the electrochemical potential of both right and left-going electrons imposed by the reservoirs.
\label{zeromodes}

The field $\mu(x)$ contains indeed the contribution of the kinetic Hamiltonian, $v_F\rho(x)$ and that of the interaction part, besides the potential $V(x)$ \cite{ines_epj}. Using the quadratic Hamiltonian expression in Eq.(\ref{H0}) it is given by:
\begin{equation}\mu(x)=\frac{v(x)}{g(x)} \rho(x)+\int dy U(x,y)\rho(y)+eV(x).
\end{equation}

The kinetic part is included in $v/g \rho$. The first two terms on the r. h. s. can be thought as nonlocal electrochemical capacitance convoluted with the density. In case $U=0$, its is simply given by $v/g$, related to the compressibility of the TLL \cite{ines_ann,ines_epj,blanter,ponomarenko_features,egger_screening}.

In the stationary regime, we can view this equation as follow. The solution for the density $\rho_0(x)$ to this equation with $\mu(x)=\mu_0$, adjust in order to compensate $V(x)$ such that $\mu_+(x)$ and $\mu_-(x)$ are uniform, imposed by $\mu_1$,$\mu_2$. This adjustment enters precisely as a zero mode contribution $\phi_0(x)$ to the bosonic field $\Phi$ through $\rho_0=-\partial_x\phi_0/\pi$. Thus $\phi_0$ is a solution of the following equation:
\begin{equation}\label{equationphi0}
\frac{v(x)}{g(x)} \partial_x\phi_0(x)+\int dy U(x,y)\partial_y\phi_0(y)=\frac h 2\omega(x),
\end{equation}
where the r. h. s is an important frequency we use throughout this paper:
\begin{eqnarray}\label{omega(x)}
\omega(x)&=&\frac{e}{\hbar}\left[\frac{V_1+V_2}2-V(x)\right].
\end{eqnarray}
We are not intending to solve the equation for $\phi_0$, depending on $v,g,U$. Let us write its simple solution in case $U=0$ \cite{ines_ann,ines_prb_long,ponomarenko_features,dolcini_stm}:
\begin{equation}
\frac {v(x)}{g(x)}\partial_x\phi_0(x)=\frac h 2\omega(x)
\end{equation}
It can be solved, letting $|x|<L/2$:
\begin{equation}\label{phi0short}
\phi_0(x)=\frac h 4 \int_{-L/2}^{L/2} sgn(x-y) \frac{g(y)}{v(y)}\omega(y).
\end{equation}
We need to find as well the time-independent zero-mode part $\theta_0(x)$ which adds to $\Theta(x)$. Using Eq.(\ref{juniversal}) and Eq.(\ref{H0}), one has the current expression:
\begin{equation}\label{rhoj}
j(x)=\frac 1 {\pi}\partial_x\Theta(x).
\end{equation}
 Thus $\partial_x\theta_0(x)=\pi I_0/v(x)g(x)$. It is convenient to introduce the frequency:
 \begin{equation}\label{omega120}
 \omega_{12}= eI_0,
 \end{equation}
 such that
 \begin{equation}\label{theta0}
 \theta_0(x)= \int_{-L/2}^{L/2} sgn(x-y) \frac{1}{v(y)g(y)}\omega_{12}.
 \end{equation}
This is a dual to $\phi_0(x)$ when $U=0$, but holds for any finite $U$ as well.
To summarize, we need to perform translation of the bosonic fields $\Phi$ and $\Theta$ by zero modes contributions:
\begin{eqnarray}\label{translation}
\Phi&\rightarrow &\Phi +\omega_{12} t+\phi_0(x)\nonumber\\\Theta &\rightarrow &\Theta +\omega(x)t+\theta_0(x).
\end{eqnarray}
where $\omega_{12}$ given by  Eq.(\ref{omega120}), and $\omega(x)$ by Eq.(\ref{omega(x)}).
 Thus one has:
\begin{equation}\label{translationH}
\mathcal{H}_{W}(\Phi,\Theta)+\mathcal{H}_V\rightarrow \mathcal{H }_W\left(\Phi+\omega_{12} t+\phi_0(x),\Theta+\omega(x)t+\theta_0(x)\right),
\end{equation}
and averages in the presence of this time-dependent Hamiltonian will be denoted by $<..>_W$. Contributions to the quadratic part have to be taken into account.
The same procedure will be used when the injector is present, whose voltage $V_3$ will be taken into account through the potential $V(x)$ in Eq.(\ref{coupling}). The profile of $V(x)$, which enters in Eq.(\ref{omega(x)}) will be determined now in the presence of a gate, and will be given by Eq.(\ref{V(x)}) with an injector.
 \subsubsection{The potential $V(x)$ in the presence of a gate and two reservoirs}
\label{chemicalgate}

The screening by the gate is often treated by a semi-classical approach. Here we propose a microscopic way similar to that in section \ref{invasive}. We express the voltage $V(x)$ entering in Eq.(\ref{coupling}) in the presence of the gate and in the two-terminal geometry.
  Let us first notice that we include implicitly in $V(x)$ internal contributions due to forward processes in $\mathcal{H}_{nq}$ \cite{ines_ann}. Contrary to external voltages which can be for instance controlled externally and can depend on time, those contributions can be rather viewed as an inhomogeneous change of the zero-modes, or a local effective potential. One of those contributions is that due to the reduction of some of the non-quadratic interaction terms to quadratic ones \cite{ines_ann}, which reads:
 \begin{equation}\label{deltaV(x)}
  \delta V(x)=\int \frac{dy k_F^2}{\pi} y\bar{U}(x,y) \sin 2k_Fy
  \end{equation}
  where we denote for any function $F$ by
  \begin{equation}\label{barF}
  \bar{F}(x,y)= F\left(x+\frac y 2,x-\frac y 2\right).
  \end{equation}
 We saw in \ref{invasive} that a similar correction occurs due to interactions with the injector, Eq.(\ref{deltaV(x)inj}). Another simpler case is that of an impurity potential $\lambda(x)$ which contributes to $V(x)$ due to the forward scattering.

 Let us now include the gate voltage. For this, we write the Coulomb interactions between the wire and the gate:

 \begin{equation}\label{Wgate}
   \mathcal{H}_{W,gate} = \int_{W} dx \int_{gate}d{{\vec{r}}}
U_{Coul}(x,{\vec{r}}) \rho(x)\rho_{gate}({\vec{r}}).
\end{equation}
 We retain only the long-wavelength part of the density in the wire, as we don't expect the form gate to have abrupt variations on the scale of $\lambda_F$. The fluctuating density part of the gate can be integrated out and induce screening effects, see \ref{invasive}. Here we consider only the average excess density in the gate,  given by:
 \begin{equation}
 <\rho_{gate}>=v_{F,gate} \mu_G,
 \end{equation}
 which yields a coupling of the density in the wire to the effective potential:
\begin{equation}\label{V(x)gate}
V_{gate}(x)
=v_{F,G} \mu_G\int_{gate}
U_{Coul}(x,{\vec{r}})d{{\vec{r}}}.\end{equation}
  This is an expression of the electronic capacitance.
Thus inside the wire, the potential $V(x)$ is given by:
 \begin{equation}\label{V(x)two}
 V(x)=V_{gate}(x)+\delta V(x)+\lambda(x),
 \end{equation}
  where $\lambda(x)$ is a possible impurity potential. We have to recall that it can contain other terms depending on the nature of the processes in play.

\subsection{Formal expression of the differential conductance and the noise}

 The results obtained in \ref{zeromodes} were based on the boundary conditions (\ref{boundaryconditions}) and concepts introduced in Ref.(\cite{ines_epj}). They allow to obtain in a universal way the zero modes, whatever the quadratic part is, and to clarify the origin of the frequency (\ref{omega(x)}). In the following, we will rather include leads with uniform potential as in Eqs.(\ref{coupling},\ref{profile})\cite{trauzettel_04,dolcini_05,ines_ff_noise,dolcini_stm}. We could have obtained the same translation in Eq.(\ref{translation}). We don't give details, but mention two crucial results allowing to recover those translations independent on the profile of interactions for their time-dependent parts. This is because they will be useful for the current correlations as well.

 One is that of the DC conductance.
 Indeed, in view of Eqs.(\ref{rhoj},\ref{coupling}), the differential non-local conductivity is given by:
\begin{equation}
\label{conductivity}
 {\sigma}(x,y,\omega)=2\omega\frac{e^2}{\pi h}
C_{\Phi,\Phi}(x,y,\omega),
\end{equation}
where we denote by
\begin{equation}\label{defC}
C_{X,Y}(x,y,t-t')
 =\theta(t-t')\left<\left[X(x,t),Y(x',t')\right]\right>
 \end{equation} for any two operators $X,Y$.
  The two-terminal DC conductance is:
\begin{equation}\label{G12}
G_{12}= {\sigma}(x,y,\omega=0).
\end{equation}
 We have seen that when the Hamiltonian of the wire reduces to $\mathcal{H}_q$ (Eq.(\ref{H0})), $G_{12}=G_{12}^q$ is given simply by Eq.(\ref{G12q}). Notice that in this case, the current becomes exactly linear in the voltage\cite{ines_schulz}. Thus the first useful result is: \begin{equation}
lim_{\omega\rightarrow 0}[2\omega C^q_{\Phi,\Phi}(x,y,\omega)]=2 g_L\label{limitCphiphi},
\end{equation} determined only by the asymptotic value $g_L$\cite{ines_bis,annales,ines_epj}. We add the superscript $q$ to recall that this is in the presence of $\mathcal{H}_q$ only. In particular, the frequency $\omega_{12}$ in Eq.(\ref{omega120}) is still given by $I_0$, which contains now $g_L$:
\begin{equation}
\omega_{12}=\frac{eg_L}{2\hbar}(V_{1}-V_{2})=\frac{eg_L}{2\hbar}V_{12}\label{omega12}.
\end{equation}
 Another Green's function which will be of crucial use is that between the fields $\Phi$ and $\Theta$, Eq.(\ref{defC}). It can be shown to be universal in the zero-frequency limit, similar to their commutator:
 \begin{equation}\label{Cmixedzero}
 C^q_{\Phi,\Theta}(x,y,\omega=0)=i\frac{\pi}2 sgn(x-y).
 \end{equation}

We can already use Eq.(\ref{limitCphiphi}) to express the DC conductance (see Eq.(\ref{G12})) formally in the presence of the total Hamiltonian $\mathcal{H}_W$, Eq.(\ref{Hdecomposed}). One can show that the differential AC conductivity, Eq.(\ref{conductivity}), obeys a Dyson equation \cite{ines_ann,ines_schulz,ines_ff_noise}:
 \begin{equation}\label{dyson}
 \sigma(x,y,\omega)=\sigma^q(x,y,\omega)+\int dx'dy' \sigma^q(x,x',\omega)\sigma^{nq}(x',y',\omega)\sigma^q(y',y,\omega),
 \end{equation}
 where $\sigma^q$ is the conductivity in the presence of $\mathcal{H}_q$ only, and:
 \begin{equation}\label{sigmanq}
\sigma^{nq}(x',y',\omega) =\frac{1}{\hbar \omega} \int_0^\infty dt \left( e^{i \omega t}-1 \right)
\left\langle \left[ j_{nq}(x',t),j_{nq}(y',0) \right]
\right\rangle_W.
\end{equation}
Let us precise that Eq.(\ref{dyson}) is valid for any voltages and temperatures, thus for the non-linear regime as well. Here we have introduced the generalized forces:
\begin{equation}
j_{nq}(x,t) =  -\frac{e }{h}\frac{\delta
\mathcal{H}_{nq}(\Phi+\phi_0(x)+\omega_{12}t)}{\delta \Phi(x,t)}  \label{jnq} .
\end{equation}
In the stationary regime, the total average current:
\begin{equation}\label{Inq}
I_{nq}=\int dx \left<j_{nq}(x,t)\right >_W
\end{equation}
 is time-independent, and its differential is:
 \begin{equation}\label{Gnq12}
 G^{nq}_{12}=\frac{d I_{nq}}{dV_{12}}=\int dx' dy' \sigma^{nq}(x',y',0),
 \end{equation}
 where $\sigma^{nq}$  becomes a derivative at zero frequency of the retarded Green's function for $j_{nq}$, Eq.(\ref{defC}). The DC differential conductance can be expressed, in view of Eqs.(\ref{G12},\ref{G12q}), as:
 \begin{equation}\label{G12nq}
 G_{12}=\frac{e^2g_L}h\left[1-g_LG^{nq}_{12}\right].
 \end{equation}

 Let us now consider the noise auto- or cross-correlations:

 \begin{eqnarray}\label{noisetwo}
S\left(\eta,\eta'\right)&=&\eta\eta'\int dt  \left<\Delta j\left(\frac \eta L 2,t\right)\Delta j\left(\eta'\frac L 2,0\right)\right>,
\end{eqnarray}
where $\eta,\eta'=\pm $ correspond to the contacts at $-L/2$ and $L/2$, and $\Delta j=j-<j>$.
 By current conservation, and as we have taken the zero-frequency limit, one has simply $S(+,+)=S(-,-)=-S(+,-)=-S(-,+)=S$, which can be expressed formally at any temperature and voltage $V_{12}=V_1-V_2$ (see Eq.(\ref{omega12})):
  \begin{equation}\label{S12T}
  S=4k_BTg_L+4k_BTg_L\frac {dI_{nq}}{dV_{12}}+s_{nq,nq},
 \end{equation}
  where $s_{nq,nq}$ is the noise associated to $j_{nq}$, Eq.(\ref{jnq}):
\begin{eqnarray}\label{snqnq}
s_{nq,nq}&=&\int dx dy dt   \left<\Delta j_{nq}(x,t)\Delta j_{nq}(y,0)\right>_W.
\end{eqnarray}
In the limit $T\ll g_LV_{12}$ one gets simply:
 \begin{equation}\label{S12two}
 S(T\ll \omega_{12})=s_{nq,nq}.
 \end{equation}

 \subsection{Expansion of the DC differential conductance for an extended potential $\lambda(x)$}
 \label{subcon}
  Let us now specify  to the case where $\mathcal{H}_{nq}$ describes merely backscattering on a potential $\lambda(x)$, which couples to the total density in Eq.(\ref{rhotot}). In general, the $m=1$ term, thus backscattering with momentum loss $2k_F$, is the most dominant in the RG sense, and will be retained for simplicity in the following discussion.
 \begin{equation}\label{HB}
 \mathcal{H}_{B}=\int \frac{dx}{2i\pi }\lambda'(x) e^{2i(\Phi(x)+\phi_0(x)+k_Fx+\omega_{12}t)}+H. C..
\end{equation}
Recall that $\phi_0(x)$ depends as well on $\lambda(y)$, see Eqs.(\ref{V(x)gate},\ref{equationphi0}).  Nevertheless, even for arbitrary range and profile of interactions in $\mathcal{H}_q$, and for weak enough $\lambda'(x)$, we can formally expand the DC conductance in terms of the renormalized $\lambda'(x)$ we call $\lambda'^*(x,\omega^*)$, Eq.(\ref{omega*}).

   In the expansion of the differential DC conductance in Eqs.(\ref{G12nq},\ref{Gnq12}) come into play fermionic correlation functions, more precisely between exponentials of $\Phi$ in (\ref{HB}). These turn out to decay exponentially at a spatial separation greater than the thermal length $L_T\simeq v/T$, where $v$ can be roughly viewed as an average velocity of the plasmons. This makes it possible to expand both $G_{12}^{nq}$ and $s_{nq,nq}$ in powers of $\lambda'^*(x,\omega^*)$ in two typical situations: $L\gg L_T$ and $L\ll L_T$. We denote the typical series due to the expansion by $A_{12}$, such that:

 \begin{eqnarray} \label{vois} G_{12}^{nq}&\rightarrow & \frac {e^2} h A_{12}\nonumber\\
   s_{nq,nq}&\rightarrow & 2\pi e\omega_{12} A_{12}.
   \end{eqnarray}
 Normally the coefficients of the series are not identical in the conductance and the noise, but as we are not computing them, only the energy dependence matters here.

   Firstly, if the wire length $L\ll L_T=v/T$, one has $\omega^*=max(\omega_L,\omega_{12})$, where $\omega_L=v/L$, and we can show that:
 \begin{equation}\label{coherent}
 A_{12}=\sum_{n=0}^{\infty}a_n\left|\int dx\lambda'^*(x,\omega^*)e^{2i[\phi_0(x)+k_F x]}\right|^{2n}.
 \end{equation}
 This corresponds to a coherent limit, where in particular interference terms at $e^{2i(k_Fx-k_Fx')}$ are important.

Secondly, if we assume $L\gg L_T$, and that one can separate the support of $\lambda(x)$ into segments $D$ of length less than $L_T$, separated by distances greater than $L_T$, one gets:
 \begin{equation}\label{incoherent}
A_{12}=\sum_{n=0}^{\infty}\sum_{n,D} a_{n,D}\left|\int_{D} dx\lambda'^*(x,\omega^*)e^{2i[\phi_0(x)+k_F x]}\right|^{2n},
 \end{equation}
 where $a_{n,D}$ are coefficients depending on interactions and the wire segment, and here $\omega^*=max(k_BT/\hbar,\omega_{12})/\Lambda$.
 Notice the linear regime with respect to $V_{12}$ is reached not only at low bias compared to temperature, i. e. when $eV_{12}\ll k_BT$, but as well if $\omega_{12}=eV_{12}/\hbar\ll \omega_L$; this is because the noninteracting leads impose their "power" dependence on voltage in this limit. This is in accordance with Eq.(\ref{coherent}) (resp. (\ref{incoherent})), where $\omega^*$ is equal to $\omega_L$ (resp. $T$) when $\omega_{12}>\omega_L$ (respec. $k_BT>\hbar\omega_{12}$), thus $G_{12}$ does not depend on $V_{12}$.
 \subsection{Local backscattering}
  The above expressions apply of course if $\lambda'(x)$ is localized around only few points. In particular, for a single impurity located at $x_0$, the extension of $\lambda'(x)$ is now few wavelengths around $x_0$ and the backscattering Hamiltonian (\ref{HB}) reduces to:
 \begin{equation}\label{HBlocal}
 \mathcal{H}_{B}=\frac{k_F}{\pi}\lambda e^{2i(\Phi(x_0)+\omega_{12}t)}+ h.c.,
 \end{equation}
 where $\lambda=\int dx \lambda(x) e^{2i[\phi_0(x)+k_F x]}$ depends on voltages as well.
  The RG equations for any profile of $g(x)$ and $U(x,y)$  were derived in \cite{ines_ann,ines_prb_long}. In this case, the renormalization of the interactions by $\lambda'$ becomes negligible.

  In the limit of an infinite wire, only $V_{12}=2\hbar\omega_{12}/e$ and temperature come into play. This corresponds to $L\gg L_T$, thus one can apply Eq.(\ref{incoherent}) with a unique domain D centered around $x_0$. We can show that such an expansion has a more general form when one has not necessarily $T\gg$ neither $\ll \omega_{12}$, but have an arbitrary ratio:
 \begin{equation}\label{scaling12}
  A_{12}\simeq \sum_{n=1}^{\infty} |\lambda^*(T)|^{2n}   f_n\left(\frac{\omega_{12}}T\right).
 \end{equation}
 This behavior generalizes the known scaling law obtained at $n=1$ to all orders as well as to inhomogeneous interactions of arbitrary type. Recall that the series enters in the two-terminal differential conductance and noise through Eq.(\ref{vois}). For either $\omega_{12}\gg $ or $\ll T$, we can check that this fits with Eq.(\ref{incoherent}) as now one has  $\omega^*=max(\omega_{12},T)$.
 Similarly, one can show that in the SBS regime, thus at $\omega^* <\omega_B^*$, one has:
 \begin{equation}\label{scalingSBS}
  A_{12}\simeq \sum_{n=1}^{\infty} |t_{12}^*(T)|^{2n}   g_n\left(\frac{\omega_{12}}T\right).
 \end{equation}
Here $t_{12}^*$ is the renormalized tunneling amplitude between the two ends. It is given by Eq.(\ref{t12*}) for a uniform TLL.
  Even though we cannot explicate the coefficients of those series, they generalize those derived exactly in the uniform short-range interacting model\cite{fendley} to inhomogeneous interactions with arbitrary range and form.


  \section{Formal exact currents correlations for the wire coupled to an injector}
 \label{formal}
 Here the 1-D system is coupled by extended tunneling to an injector at an electrochemical potential $\mu_3=e V_3$, as in Fig.(\ref{fig3}). In order to write the tunneling Hamiltonian, we need to implement the potential differences between the wire and the injector. Our strategy is to take into account $\mu_3$ through the potential $V(x)$ inside the wire which enters again as in Eq.(\ref{coupling}), as we explain in subsection \ref{chemicalinj}. The way this coupling is treated is the same as that for the two-terminal geometry, explicated in \ref{zeromodes}. The translation by the zero modes is then implemented in the tunneling Hamiltonian in \ref{subHT}. We express rigorously the average currents at the three terminals in subsection \ref{subformal} and the current correlations in subsection \ref{subcrossformal}.
 \subsection{Determination of the potential $V(x)$ inside the wire}

 \label{chemicalinj}
 In \ref{chemicalgate}, we showed how Coulomb interactions with the gate as well as interactions inside the wire change the profile of $V(x)$ in Eq.(\ref{V(x)gate},\ref{V(x)two}). Here we add to Eq.(\ref{V(x)two}) the effect of the injector. Nevertheless, we don't
determine the potential through any self-consistent argument. The electronic
interactions are included in the Hamiltonian, and are taken as well into account between the injector and the wire, besides those between the gate and the wire as analyzed in \ref{chemicalgate}.
 The
three-dimensional potential $V(\vec{r})$ obeys the
boundary conditions imposed by the four terminals with fixed
electrochemical potentials: the injector, the two reservoirs and the gate. Let us include the Coulomb interactions between the wire (labeled by W) and the injector:
  \begin{equation}\label{Winj0}
   \mathcal{H}_{W,inj} = \int_{W} dx \int_{inj}d{{\vec{r}}}
U_{Coul}(x,{\vec{r}}) \rho_{tot}(x)\rho_{inj}({\vec{r}}),
\end{equation}
where $\rho_{tot}(x)$ is given by Eq.(\ref{rhotot}) and $\rho_{inj}$ is the density in the injector, \begin{equation}\label{rhoinj0}\rho_{inj}(\vec{r})= \left<\rho_{inj}\right>+\delta\rho_{inj}(\vec{r}),\end{equation}containing, besides the density fluctuations $\delta\rho_{inj}$, the average value of the excess density. We
retain here only the role of the latter,
\begin{equation}<\rho_{inj}>\simeq  v_F \mu_3,\end{equation}
postponing that of the fluctuations to section \ref{invasive}. Its contribution to Eq.(\ref{Winj}) reads:
\begin{equation}\label{total}
\int V_{inj}(x)\rho_{tot}(x),
\end{equation}
where
\begin{equation}\label{Vinj0}
V_{inj}(x)=v_{F,inj} \mu_3 \int_{inj}
U_{Coul}(x,{\vec{r}})d{{\vec{r}}}.
\end{equation}
The density fluctuations in the injector and the coupling to the higher Harmonics in Eq.(\ref{total}) will be commented when invasive effects are analyzed in section \ref{invasive}. The
 coupling to the long-wavelength part of the density in the wire $\rho$ adds a contribution $V_{inj}(x)$ to eq.(\ref{V(x)two}). There is as well another correction $\delta V_{inj}$ which will be shown to arise from the integration of the  density fluctuations in the injector, see Eq.(\ref{deltaV(x)inj}).
  Voltages are implicitly defined up to a global constant, which ensures gauge invariance. Indeed $V(x)$ is incorporated through a coupling to the density in the wire. Through the form adopted in Eq.(\ref{rhotot}), a global translation of $V(x)$ by a constant does not have any effect.
Now we can write the total effective potential inside the wire:
\begin{equation}\label{V(x)}
V(x)=V_{inj}(x)+V_{gate}(x)+\delta V_{inj}(x)+\delta V(x)+\lambda(x)  for |x|<L/2
\end{equation}
where the different contributions are written in Eqs.(\ref{Vinj},\ref{V(x)gate},\ref{deltaV(x)inj},\ref{deltaV(x)}).

 $V(x)$ enters in Eq.(\ref{omega(x)}), which itself controls both time and space dependence in the zero modes, see Eqs.(\ref{phitheta},\ref{equationphi0},\ref{theta0},\ref{translation}). Nevertheless, the time-dependent zero mode of the field $\Phi$ in Eq.(\ref{translation}) does not change as it depends only on the difference between the two voltages of the reservoirs $=V_1-V_2$, or $\omega_{12}$, Eq.(\ref{omega12}). This is a crucial point to stress here.
\subsection{Tunneling Hamiltonian in an out-of equilibrium situation}
\label{subHT}
 Let us now adopt tunneling with arbitrary spatial extension between the injector and the wire, described by a Hamiltonian $\mathcal{H}_T$.  Taking the translations in Eqs.(\ref{translation},\ref{translationH}) into account, one can write the
 total Hamiltonian, adding $\mathcal{H}_T$ and the Hamiltonian of the injector $\mathcal{H}_{inj}$ to $\mathcal{H}_W$, Eq.(\ref{Hdecomposed}):
\begin{equation}
{\mathcal{H}} ={\mathcal{H}}_q  \, + \, {\mathcal{H}}_{nq}  \,
+{\mathcal{H}}_T+\mathcal{H}_{inj}. \label{Htot}
\end{equation}
 We will also show that invasive effects (see section\ref{invasive}) induce inhomogeneous corrections to all types of terms in Eq.(\ref{Hdecomposed}) and Eq.(\ref{H0}). As we will not specify the Hamiltonian for the present formalism, such corrections are implicitly included. In the following, we assume the Hamiltonian $\mathcal{H}_{nq}$ to depend only on $\Phi$, with an exponential non-local dependence. This is a generic situation in view of Eq.(\ref{rhotot}). In particular other terms can in general be reduced to linear or quadratic forms, see \ref{chemicalgate}.

 The tunneling Hamiltonian is written in terms of the right and left fermion field operators, Eq.(\ref{Psir}) for $r=\pm$:
 \begin{equation}\label{HT}
{\mathcal{H}}_{\mathrm{T}}=\sum_{r=\pm} \int_W dx\int_{inj} d\vec{r}\Gamma_r(x,\vec{r})
\Psi_{r}^{\dagger}(x)\Psi_{inj}(\vec{r}) e^{i \omega_r(x) t}+\mathrm{h.c.},
\end{equation} where $\Psi_r$ is given by Eq.(\ref{Psir}). One could as well have distinguished the fields in the injector coupling to $\Psi_r$, which would make expressions more cumbersome.
We have chosen to take the zero-modes out from $\Psi_r$. On one side, it is implicit that:
 \begin{equation}
 \Gamma_{r}\rightarrow \Gamma_{r} e^{ir k_F x+ir\phi_0(x)-i\theta_0(x)},
 \end{equation}
 where $\phi_0(x)$ obeys Eq.(\ref{equationphi0}), and is given by Eq.(\ref{phi0short}) for short-range interactions. $\theta_0(x)$ is given by Eq.(\ref{theta0}). Thus even when the bare amplitudes are originally equal for $r=\pm$, they become different out-of-equilibrium. On the other side, the translation in Eqs.(\ref{phitheta},\ref{translation}) gives (see also Eq.(\ref{thetwo}):
 \begin{eqnarray}
\omega_r(x)&=&\frac{e}{\hbar}V_r(x)=\omega(x)+r\omega_{12},\nonumber
\\V_r(x)&=& \frac{rg_L+1}2 V_2-\frac{rg_L-1}2 V_1-V(x),\label{Vr}
\end{eqnarray}
 where $V(x)$ is given by Eq.(\ref{V(x)}). For $g_L=1$, thus in the presence of noninteracting leads, one has simply $V_+(x)=V_2-V(x)$ and $V_-(x)=V_1-V(x)$. Notice that usually, taking $V_1=V_2$, one implements by hand $V_+=V_-=V_2-V_3$ instead. Here the voltages for right and left-going electrons tunneling are different, screening of $V_3$ is taken into account.
Let us comment on $V_r $ in Eq.(\ref{Vr}), which intervenes in the zero modes for the bosonic fields $\Phi_r$ for right ($r=+$) and left ($r=-$)-going electrons. The way they are expressed in terms of $V_1$ and $V_2$ is similar to that of the chiral right and left propagating density modes in Eq.(\ref{Phichiral}) \cite{ines_schulz,ines_ann,pham}.
 Here they can be defined so simply only where interactions are short-range with constant interaction parameter, and with $\mathcal{H}_{nq}=0$, here on the asymptotic "leads" with parameter $g_L$. We see that in Eq.(\ref{Vr}), it is $g_L$ which intervenes, independently of the internal interaction profile and range. This holds in particular for uniform $g$ inside the wire: an incident right-going soliton charge reaching the contact gets reflected back with a coefficient $-\gamma$, Eq.(\ref{gamma}). Its multiple reflection process continue on to affect Eq.(\ref{Vr}), replacing $g$ by $g_L$. Recall that we could have obtained the zero modes by absorbing the coupling to $V(x)$, using Eq.(\ref{limitCphiphi}). The factor $g_L$ comes precisely from the DC conductance $G_{12}^q$ in Eq.(\ref{G12q}) whose value, in turn, is due to the same multiple reflection process when an electron is incident from the leads. $G_{12}^q$ was shown to be exactly given by $g_L$ multiplied by the transmission through the wire which becomes perfect in the zero frequency limit whatever the type of interactions is \cite{ines_ann,ines_epj}.

\subsection{Exact formal expressions for the out-of equilibrium currents at the three terminals}
\label{subformal}
We give here formal expressions for currents at both contacts, given by the average of the operators $j(\pm L/2)$ in Eq.(\ref{rhoj}) with the Hamiltonian $\mathcal{H}_W$, Eq.(\ref{Htot}), in particular for arbitrary profile of the tunneling amplitudes. They are indeed valid more generally for any Hamiltonian $\mathcal{H}_T$ which depends on both $\Phi,\Theta$.

 We could as well have expressed the AC differential conductance matrix. It has now many elements as one has three potential values $V_l$ for $l=1..3$, besides the gate voltage. They obey generalized Dyson equations at finite frequency similar but more involved than Eq.(\ref{dyson}). They imply bosonic Green's functions between $\Phi$ and $\Theta$ in the presence of $\mathcal{H}_q$ only, Eq.(\ref{H0}) and whose values are universal in the zero-frequency limit, given by Eq.(\ref{limitCphiphi}) and Eq.(\ref{Cmixedzero}).
We prefer to give here only the DC nonlinear currents expressions for arbitrary voltages $V_l$ and temperatures. Their form does not depend on the type of interactions inside the wire in Eq.(\ref{H0}). This is due precisely to the universal values in Eqs.(\ref{limitCphiphi},\ref{Cmixedzero}). The currents at the terminals $1$ thus at $-L/2$ and at $2$, i. e. $-L/2$ are:

  \begin{equation}\label{I12}
  I(\pm L/2)=\pm\left< j\left(\pm\frac L2\right)\right>= \pm [I_0-
 g_LI_{nq}^T]+\sum_{r=\pm}\frac {\pm r g_L+ 1}2 I_{r},
  \end{equation}
  where \begin{equation}\label{I0}
  I_0=\frac{e^2}hg_L(V_2-V_1),
  \end{equation}
  and:
 \begin{equation}\label{Ir}I_X=\int dy <j_X(y,t)>,\end{equation} for $X=\pm,nq$ are time-independent. Here the generalized forces $j_r$ are defined by:
\begin{eqnarray}
j_r({{x}},t)&=&-\frac e{h}\frac{\delta
\mathcal{H}_{T}(\Phi_{r}+\omega_r(x) t)}{\delta\Phi_r({x})}.\label{jr}
 \end{eqnarray}
$j_{nq}$ has the same definition as in Eq.(\ref{jnq}) in the absence of tunneling; nevertheless its average is now different from Eq.(\ref{Inq}) as it is computed in the presence of $\mathcal{H}_T$, which motivates the superscript T on $I_{nq}$. Notice that we have taken the convention of outgoing currents, justifying the added minus sign on the left contact. Kirchoff's law is obeyed, $\sum_{l=1}^3 I_l=0$ where $I_3$ is the total current in the injector given by:
\begin{equation}\label{IT}
I_3=-I_T=-\sum_{r=\pm} I_{r}.
\end{equation}

Let us also draw attention to the last term on the r. h. s. of Eq.(\ref{I12}) which is, similarly to Eq.(\ref{Vr}), related to the chiral modes' expression, Eq.(\ref{Phichiral}). In particular, it reproduces the phenomena mentioned before: a current $I_{r}$ injected inside the wire in direction $r=\pm$ "sends" a current proportional to $(rg_L+1)/2$ to the right contact and $(-rg_L+1)$ to the left contact. Nevertheless, those are controlled by the parameter of the leads $g_L$ independently of the interactions in Eq.(\ref{H0}).

 Let us now comment briefly the case when $\mathcal{H}_{nq}=0$, letting $V_1\neq V_2$. Then Eq.(\ref{I12}) simplifies to:
\begin{equation}\label{Iq}
I(\eta L/2)= \eta I_0
 +\sum_{r=\pm}\frac {\eta r g_L+ 1}2 I_{r}.
\end{equation}
If one specifies further to the realistic situation with noninteracting leads, when $g_L=1$, $V_+=V_2-V(x)$ and $V_-=V_1-V(x)$, Eq.(\ref{Iq}) has a form similar to that in a noninteracting wire:
\begin{equation}\label{Iq1}
 I(\pm L/2)=\pm\frac{e^2}h(V_2-V_1)+ I_{\pm},
\end{equation}
thus the current at the right (left) contact is determined by the total right (left)-going tunneling current average, Eqs.(\ref{jr},\ref{Ir}). One consequence is that if one injects only left-going electrons, and for $V_1=V_2$, one has $I(-L/2)=I_-$ and $I(L/2)=0$, i. e. no asymmetry. This result corresponds to that in Ref.\cite{yacoby_theo} for perfect contacts, as well as in the work \cite{dolcini_stm} which was indeed not in contradiction with the former, and from which no additional conclusions on the interacting case can be drawn.

 \subsection{Formal exact expressions for the current auto- and cross-correlations}
 \label{subcrossformal}
 Now consider noise correlations, defined as in Eq.(\ref{noisetwo}) but in the presence of the injector, for $\eta,\eta'=\pm$ by:
\begin{eqnarray}\label{defcross}
S\left(\eta,\eta'\right)&=&\eta\eta'\int dt  \left<\Delta j\left(\frac \eta L 2,t\right)\Delta j\left(\eta'\frac L 2,0\right)\right>,
\end{eqnarray}
where $\Delta j=j-<j>$, see Eq.(\ref{rhoj}). Thus $\eta=\eta'$ correspond to auto-correlations at the same contact $\eta L/2$, while $\eta'=-\eta$ correspond to cross-correlations.
For the general Hamiltonian Eq.(\ref{Htot})
we can express them in terms of the correlations defined for the labels $X,Y=+,-,nq$
\begin{equation}\label{correlations}
s_{X,Y}=\int dx dy dt   \left<\Delta j_X(x,t)\Delta j_Y(y,0)\right>
\end{equation}
where $j_{r}$ for $r=\pm$ is defined in Eq.(\ref{jr}) and $j_{nq}$ in Eq.(\ref{jnq}). We add a subscript $T$ to $s_{nq,nq}$ to distinguish it from the two-terminal noise $s_{nq,nq}$ in the presence of the Hamiltonian of the wire only, Eq.(\ref{snqnq}).
One can show the exact expression:
\begin{equation}\label{S12}
4S(\eta,\eta')=-4g_L^2\eta\eta's_{nq,nq}^T+\sum_{r,r'}(1+r\eta g_L)(1+r'\eta'g_L)s_{r,r'}-g_L\sum_r (1+r\eta g_L)s_{r,nq}+(1+r'\eta'g_L)s_{nq,r},
\end{equation}
up to a contribution which vanishes at zero temperature, and which we will express elsewhere because of its great relevance. This is not required for our main purpose here. Equation (\ref{S12}) is a new expression for a three out-of-equilibrium wire geometry, and applies for instance in crossed or parallel interacting wires. Notice the appearance again of $1+\eta r g_L$ which multiplies effectively each current $j_r$, and expresses the partial charge soliton reaching the contact $\eta L/2$ when a current $j_r$ is injected. The current $j_{nq}$ is multiplied by $\eta g_L$, as $g_L$ comes from Eq.(\ref{limitCphiphi}).

We must also notice that contrary to the two-terminal setup, one has not necessarily $S(+,+)=S(-,-)$ neither $S(+,-)=S(-,+)$. One can however symmetrize the cross correlations to get:
\begin{eqnarray}\label{symmetrizedcross}
S_{sym}(+,-)&=&\frac{1} 2\left[S(+,-)+S(-,+)\right]\nonumber\\
&&=-g_L^2 s_{nq,nq}+ \sum_{rr'}(1-rr'g_L^2)s_{rr'}-g_L^2\sum_r \frac r 2 \left[s_{nq,r}+s_{r,nq}\right].
\end{eqnarray}
As the total tunneling noise is given by:
\begin{equation}
S_T=\frac 1 4\sum_{r,r'}s_{r,r'},
\end{equation}
we notice that:
\begin{equation}
\sum_{\eta,\eta'=\pm}S(\eta,\eta')=S_T,
\end{equation}
which is the noise on the injector.
Recall that all those expressions are non perturbative. In particular they do not require any weak tunneling amplitude neither weak $\mathcal{H}_{nq}$, and holds for three different potentials in the injector and the two reservoirs connected to the wire with any inhomogeneous Hamiltonian (\ref{Htot}). They require only one defines asymptotic leads with a constant parameter $g_L$. It is equal to $1,v_F$ in a realistic situation, or to $0$ if a semi-infinite wire is considered. For $g_L=1$, equation (\ref{S12}) simplifies to:
\begin{equation}\label{S12gL1}
S(\eta,\eta')=\eta\eta's_{nq,nq}+s_{\eta,\eta'}-2\left[\eta s_{\eta',nq}+\eta' s_{nq,\eta}\right].
\end{equation}
In particular, in the absence of any non-quadratic Hamiltonian, this indicates that the cross-correlations between $j(L/2)$ and $j(-L/2)$ are equal to those between $j_+$ and $j_-$ respectively.
The cross-correlations will be discussed in the next section for weak extended tunneling and in section \ref{sharp} for a local tunneling.

\section{Weak tunneling and crucial results for three-terminal out-of-equilibrium transport}
\label{weaktunneling}
 This section contains crucial features which have been overlooked in three-terminal out-of-equilibrium transport. They are generally valid for any tunneling strength, and a general Hamiltonian in Eqs.(\ref{Htot},\ref{Hdecomposed}). But we specify to weak tunneling with an arbitrary spatial extension in the following, both for simplicity and for connection to many experimental setups. We also specify to the most interesting case of relevant processes in the RG sense in $\mathcal{H}_{nq}$. Recall that we consider the generic situation where this has an exponential dependence on $\Phi$ (which can be non-local). Thus $\mathcal{H}_{nq}$ contains oscillating time dependence factors $e^{im\omega_{12}t}$ in view of the translation (\ref{translation}). We express the currents averages and cross-correlations (subsection \ref{subGamma}) to second order in tunneling, without any expansion in $\mathcal{H}_{nq}$. For arbitrary different voltages in the three terminals and temperatures, this requires Keldysh formalism. Nevertheless, in order to avoid cumbersome expressions, we will only give the structure of the different terms. In subsection \ref{subzero}, we ask one crucial question relevant to the frequent situation when the wire is grounded, thus at $\omega_{12}=0$: if the voltage of the injector, or more precisely $\omega_+$ which in this limit is equal to $\omega_-$ (Eqs.(\ref{Vr},\ref{HT})) can offer a cutoff from the RG growing of $\mathcal{H}_{nq}$. The answer turns out to be negative! In order to prove in a more clear way this argument,  we discuss formally the perturbative expansion to all powers of $\mathcal{H}_{nq}$ at non-vanishing $\omega_{12}$ and finite temperature in subsection \ref{subexpansion}. This will be illustrated in the case of a sharp tip or STM in sections \ref{sharp} and \ref{scaling}. We also show the crucial non-equilibrium regime criteria, Eq.(\ref{out0}). The cross-correlations are considered again in subsection \ref{subcross}, where we will specify to this regime in order to get rid of thermal fluctuations. They are shown to be dominated by their two-terminal value, thus to be negative for local processes in $\mathcal{H}_{nq}$.

 \subsection{Current correlations: non-perturbative expressions with respect to $\mathcal{H}_{nq}$}
 \label{subGamma}
Here we expand formally the currents at the contacts, thus $I_{r}$ and $I_{nq}^T$ on the r. h. s. of Eq.(\ref{I12}), to lowest order in $\Gamma_{\pm}$, without any expansion with respect to $\mathcal{H}_{nq}$. This leads to correlation functions in the absence of tunneling, thus in the presence of the wire Hamiltonian $$\mathcal{H}_W(t)=\mathcal{H}_q(\Phi,\Theta)+\mathcal{H}_{nq}(\Phi+\phi_0(x)+\omega_{12}t)$$ which we denoted by $<..>_W$, while average with respect to the Hamiltonian of the injector $\mathcal{H}_{inj}$ is denoted by $<..>_{inj}$.

It is convenient to denote, for labels $r_1,r_2=\pm$ referring to right or left-going electrons, and $l$ to $(x_l,t_l)$ for $l=1,2$:
\begin{equation}\label{lambdar1r2}
\lambda_{r_1,r_2}(1,2)=\int_{inj} d\vec{r}_1\int_{inj}  d\vec{r_2}\Gamma_{r_1}(x_1,\vec{r}_1)\Gamma_{r_2}^{\dagger}(x_2,\vec{r}_2) G^{inj}_{ferm}(\vec{r}_1,\vec{r}_2,t_1-t_2)e^{i(\omega_{r_1}t_1-\omega_{r_2}t_2)},
\end{equation}
where the Green's functions in the injector have to be normally Keldysh ones, and are related to their imaginary time expression:
\begin{equation}\label{Gferminj}
G_{ferm}^{inj}(\vec{r}_1,\vec{r}_2,\tau_1-\tau_2)=\left< T_{\tau}\Psi_{inj}(\vec{r}_1,\tau_1)\Psi_{inj}^{\dagger}(\vec{r}_2,\tau_2)\right>_{inj}.
\end{equation}
Now the tunneling currents $I_{r}$ in Eqs.(\ref{Ir},\ref{jr}) can be written roughly as:
\begin{eqnarray}\label{IrGamma}
I_{r_1}&\simeq & \sum_{r_2=\pm}\int dx_1 dx_2 dt_2  \lambda_{r_1,r_2}(1,2) \left<\Psi_{r_1}(1)\Psi_{r_2}^{\dagger}(2)\right>_W,
\end{eqnarray}
as well as Hermitian conjugate expressions. Here we abbreviate $l=1,2$ for $(x_l,t_l)$. Notice that $\omega_{12}=\omega_+-\omega_-$, Eqs.(\ref{omega12},\ref{Vr}), thus one can show that the integrand depends only $t_1-t_2$, leading to a stationary current $I_r$.

  One can as well express $I_{nq}^T$ in Eq.(\ref{I12}) as the superposition of the two lowest-order terms with respect to tunneling: one being the exact average $I_{nq}$ in the absence of tunneling, Eq.(\ref{Inq}) (which can however be affected by the invasive effects compared to the isolated wire), thus in the presence of $\mathcal{H}_W(t)$ only, and another term to second order in tunneling we call $\delta I_{nq}^T$:
\begin{equation}\label{Idecomposed}
I_{nq}^T=I_{nq}+\delta I_{nq}^T+o(\Gamma^4).
\end{equation}
where the form of $\delta I_{nq}^T$ is given by:
\begin{eqnarray}\label{deltaInqT}
\delta I_{nq}^T&\simeq& \int  \lambda_{r_1,r_2}(1,2)\left<\Psi_{r_1}(1)\Psi_{r_2}^{\dagger}(2)j_{nq}(3)\right>_W,
\end{eqnarray}
where integration runs over the two times $t_1$ and $t_2$, and the three spatial coordinates. $\lambda_{r_1,r_2}$ is given by Eq.(\ref{lambdar1r2}). In general $j_{nq}$, Eq.(\ref{jnq}), can contain two or more fermionic operators, thus one has now at least four-point correlation functions in the presence of $\mathcal{H}_W(t)$ only.
The differential of $I_{nq}$ with respect to $\omega_{12}$ is given by the two-terminal conductance $G_{12}^{nq}$ we introduced in the first section, Eqs.(\ref{Gnq12},\ref{G12nq}). It was expanded in the special case when $\mathcal{H}_{nq}=\mathcal{H}_B$ describes backscattering on a potential $\lambda$ in Eqs.(\ref{coherent},\ref{incoherent}). In particular, for $\omega_{12}=0$, one has $I_{nq}=0$, while $G_{12}^{nq}$ becomes a function of temperature and other energy scales associated to $\mathcal{H}_W$. Nevertheless, if the differential of the currents at the contacts is taken with respect to the voltage of the injector $V_3$, the contribution of $I_{nq}$ drops, and in view of Eq.(\ref{I12}), one has:
\begin{equation}\label{differentialcurrent}
 \frac {\partial{I(\pm L/2)}}{\partial V_3}=\pm \frac{\partial{\delta I_{nq}^T}}{\partial V_3}+\frac{\partial I_T}{\partial V_3},
 \end{equation}
all depending on $\omega(x)$ and $\omega_{12}$ in Eqs.(\ref{omega(x)},\ref{omega12}). Here $I_T=\sum_r I_r$, and expressions of $I_r$ and $\delta I_{nq}$ are given by Eqs.(\ref{IrGamma},\ref{deltaInqT}).

Now we express the auto-correlations, and particularly the cross-correlations in Eq.(\ref{S12}) of interest here, to lowest order in tunneling amplitudes, using the non-local effective amplitude (\ref{lambdar1r2}). One needs the expressions of the different $s_{X,Y}$ in Eq.(\ref{correlations}) which intervene on the r. h. s. of Eq.(\ref{S12}).

 First, similarly to $I_{nq}^T$ which contains $I_{nq}$ in Eq.(\ref{Idecomposed}) and which are the average of $j_{nq}$ respectively for the Hamiltonian in Eq.(\ref{Hdecomposed}) and Eq.(\ref{Htot}), one has also :
 \begin{equation}\label{snqdecomposed}
 s_{nq,nq}^T=s_{nq,nq}+\delta s_{nq,nq}^T,
 \end{equation}
where $s_{nq,nq}$ is the exact noise in a two-terminal geometry with a bias $V_{12}=2\hbar\omega_{12}/eg_L$, in the presence of the Hamiltonian $\mathcal{H}_W$ in Eq.(\ref{Hdecomposed}). The second term is in second order of tunneling:
\begin{equation}\label{snqnqGamma}
\delta s_{nq,nq}^T\simeq\int \sum_{r_1,r_2=\pm} \lambda_{r_1,r_2}(1,2)\left<\Psi_{r_1}(1)\Psi_{r_2}^{\dagger}(2)j_{nq}(3)j_{nq}(4)\right>_W.
\end{equation}
Integration runs over the four space coordinates but only the first three times.
There is as well the correlation between $j_{nq}$ and the tunneling current for right or left-going electrons $j_{r_2}$, $r_2=\pm$, which has a form similar to $\delta I_{nq}^T$ in Eq.(\ref{deltaInqT}), with $r_2$ now selected:
\begin{equation}\label{srnqGamma}
 s_{r_2,nq}\simeq \int \sum_{r_1=\pm} \lambda_{r_1,r_2}(1,2)\left<\Psi_{r_1}(1)\Psi_{r_2}^{\dagger}(2)j_{nq}(3)\right>_W.
\end{equation}
Integration runs now over $t_1,t_2$ and over all three space coordinates.
Finally one has the correlations between tunneling currents $<j_{r_1}j_{r_2}>$ whose expression is similar to the average in Eq.(\ref{IrGamma}):
\begin{equation}\label{sr1r2Gamma}
 s_{r_1,r_2}\simeq \int \lambda_{r_1,r_2}(1,2)\left<\Psi_{r_1}(1)\Psi_{r_2}^{\dagger}(2)\right>_W,
\end{equation}
 where integration runs over time $t_2$ and two space coordinates only.

  Let us give here naive expectations, and see later the subtleties related to both order of limits and to validity of perturbation with respect to tunneling. Let us specify to the case when $\mathcal{H}_{nq}=0$, thus the wire is described by a purely quadratic Hamiltonian $\mathcal{H}_q$, Eq.(\ref{H0}). In this case, Eq.(\ref{S12}) becomes, to order $\Gamma_{\pm}^2$
 \begin{equation}
 S(+,-)= (1-g_L^2)s_T.
 \end{equation} When tunneling is local, $s_T$ is positive. In the limit of zero-temperature, the second part on the r. h. s. is positive (negative) for repulsive (attractive) interactions on the leads if $g_L<1$ ($>1$). This result was found in Ref.(\cite{crepieux_stm}), where it was specified to an infinite wire with constant parameter $g_L=g$ and $V_1=V_2$. We have generalized it to an arbitrary type of interactions and for $V_1\neq V_2$, i. e. non vanishing $\omega_{12}$. We have nevertheless three additional subtleties. First, there are thermal noise contributions not written here. They contain the two first terms on the r. h. s. of eq.(\ref{S12T}) which are not necessarily negligible with respect to other terms. Second, the order of taking the limits $\omega_{12}\rightarrow 0$ and $T\rightarrow 0$ matters when interpretation in terms of quantum statistics is required. Third, there are terms of order $|\Gamma_+|^2|\Gamma_-|^2$ where the short-time contribution dominates. It corresponds to the generation of backscattering by tunneling we will show. This will yield negative cross-correlations. We have also to notice that it is not physical to measure currents where $g_L\neq 1$ asymptotically, as external voltages cannot be imposed on an interacting region.

 \subsection{Limit of $\omega_{12}=e g_L V_{12}/\hbar=0$: can the voltage of the injector offer an RG cutoff?}
\label{subzero}
An important situation which is frequently considered in both experimental and theoretical setups is that when $\omega_{12}=eg_LV_{12}/\hbar=0$. According to Eqs.(\ref{Vr},\ref{omega(x)}) this yields:
\begin{equation}\label{omegapequalomegam}
\omega_+(x)=\omega_-(x)=\omega(x),
\end{equation} which is now the effective voltage at which electrons are injected. Usually, this corresponds to the voltage tip, $$\omega(x)\rightarrow eV/\hbar$$ which we have determined here with more caution, in particular it has in general a non-trivial dependence on x, see Eq.(\ref{V(x)}). Notice that one has also $\omega_{12}=0$ if the wire is open at one end, as its DC conductance corresponds to $g_L=0$. Thus our present criticism is relevant for instance for a semi-infinite wire with an impurity investigated for instance in Ref.\cite{glazman_stm}.

In related works, it is often accepted that reaching a non-equilibrium regime is ensured by letting $V\gg T$, in which case one can take the limit of zero temperature in all correlation functions \cite{crepieux_stm,dolcini_stm}. Morever, when the injector plays the role of an STM, the DOS is deduced from expressing $I_T$ as a convolution between the DOS in the STM and that in the wire at energy $eV$.

We show now that those assertions are indeed not justified in general. Even though we illustrate it in the present model, the problems to which we are drawing attention are valid for any multi-terminal geometry.

 Let us consider the expression (\ref{Idecomposed}), where one has now $I_{nq}=0$ as there is no current in a two-terminal geometry without bias. Thus one is left with $\delta I_{nq}$ and $I_r$ whose perturbative expressions with respect to tunneling are given in Eqs.(\ref{deltaInqT},\ref{Ir}), taken now with Eq.(\ref{omegapequalomegam}) and at $\omega_{12}=0$:
 \begin{equation}\label{Iequal}
 I(\pm L/2)=\pm \delta I_{nq}^T+\sum_r I_r.
 \end{equation}
Expressions of $I_r$ and $\delta I_{nq}$, Eqs.(\ref{IrGamma},\ref{deltaInqT}) show that they correspond to Fourier transforms at frequency $\omega$ of correlation functions at equilibrium, as they are computed with respect to the Hamiltonian in Eq.(\ref{Hdecomposed}) at $\omega_{12}=0$:
 \begin{equation}\label{HWequ}
 \mathcal{H}_W=\mathcal{{{H}}}_q(\Phi,\Theta)+{\mathcal{{H}}}_{nq}(\Phi).
 \end{equation}
 We first recall briefly the frequent way of dealing with similar problems in literature. We continue to use $V$ instead of our $\omega(x)$.
 As far as equilibrium transport in its usual accepted meaning is ensured, i. e. $V\ll T$, the currents are linear with respect to $V$; the differential conductance matrix is a function of temperature and other length scales.
 The problem arises when one looks at $V\gg T$.

  As $V$ enters in the integrals of tunneling current, the Green's functions $<\Psi_{r_1}(1)\Psi_{r_2}(2)>_W$ in $I_T=\sum_{r_1}I_{r_1}$, Eq.(\ref{IrGamma}), are replaced by their zero-temperature limit. This is
  justified by the fact that the Fourier transforms of correlation functions at a frequency $V\gg T$ allows to retain the contribution of times of order $1/V$, thus much smaller with respect to $1/T$, therefore to let $T\rightarrow 0$. Nevertheless, we show here that this argument works only for a pure quadratic Hamiltonian, Eq.(\ref{H0}); in this case, the equilibrium Green's functions in Eq.(\ref{Ir}) are a function of $T(t_1-t_2)$. Only integration over this time difference intervenes, with an oscillating function $e^{iV(t_1-t_2)}$, justifying for instance Eq.(\ref{ITbulk}).
 Such procedure cannot be undertaken in the presence of $\mathcal{H}_{nq}$. We will precisely show that the two following related assertions are not appropriate:

\begin{itemize}
	\item It is stated often in literature that adopting high enough voltage $V$, i. e. taking $V$ above some typical crossover energy we call $\omega_{nq}^*$, and given by Eq.(\ref{omegaB*}) when $\mathcal{H}_{nq}$ describes one impurity, allows to make a perturbative expansion with respect to weak $\mathcal{H}_{nq}$.
	\item  In this expansion, letting $V\gg T$ allows even more to take the zero-temperature limit, as time differences very small compared to $1/T$ intervene.
\end{itemize}
Indeed one expects that the criteria of validity for the expansion is that one has at least one energy above $\omega_{nq}^*$.
 The subtle question we ask here is: which energy among the voltage combinations besides temperature allow to stop the RG flow from growing, thus to perform such a expansion? The answer could be thought to be trivial, adding simply V to the list of the IR frequency $\omega^*$ in Eq.(\ref{omega*}), as what is often assumed in the literature. But this is not so!
  \subsection{Expansion in powers of $\mathcal{H}_{nq}$ at $\omega_{12}\neq 0$}
\label{subexpansion}

In order to answer in a more clear way such a question, we need to make the formal expansion to all powers of $\mathcal{H}_{nq}$ by keeping all voltages and temperature finite, in particular letting now $\omega_{12}\neq 0$ and to find the criteria of validity afterward. We use again our $\omega(x)$ for each tunneling point $x$ instead of $V$. We will illustrate those facts for a local backscattering in section \ref{scaling}.

A first important remark is that the two-terminal current $I_{nq}$ and noise $s_{nq,nq}$ intervene already in Eqs.(\ref{Idecomposed},\ref{snqdecomposed}). Their expansion in powers of $\mathcal{H}_{nq}$ requires the same criteria as that in a two-terminal geometry, i. e. that  Eq.(\ref{omega*}) obeys:
\begin{equation}\label{criteria}\omega^*>\omega_{nq}^*.\end{equation} This might not be convincing enough as one can get rid of those two-terminal contributions either by taking differentials with respect to the voltage injector, see Eq.(\ref{differentialcurrent}), of which they are independent, or by letting $\omega_{12}=0$, in which case they vanish, see Eq.(\ref{Iequal}). Nevertheless the latter limit is precisely subtle to take as we will see.

Thus we will discuss the expansion of the other contributions to current and noise in Eqs.(\ref{IrGamma},\ref{Idecomposed},\ref{S12}), in order to prove that Eq.(\ref{criteria}) is required as well.

Each term of order $n$ in the expansion implies multiple Fourier transforms of equilibrium correlation functions between vertex fields at many times, $t_3,.. t_n$, besides $t_1,t_2$ which enter in Eqs.(\ref{IrGamma},\ref{deltaInqT}). More precisely, using the function $\lambda(1,2)$ introduced in Eq.(\ref{lambdar1r2}), the generic form of the n'th term in the expansion is:

\begin{equation}\label{formalexp}
\int e^{i\omega(x)(t_1-t_2)}g(t_1-t_2)[e^{i\omega_{12}(t_1+t_2)}\Pi_{l\neq k=1..n} e^{i\omega_{12}(t_l-t_k)} \chi(t_l-t_k)
\end{equation}
where integration runs over all times and neither spatial dependence neither amplitudes are specified, and $\omega(x)=\omega_+(x)+\omega_-(x)$, see Eq.(\ref{omega(x)}). The product runs over all pairs of times ${t_l,t_k}$ apart from ${t_1,t_2}$. As the correlation functions $\chi$ are computed with respect to the Hamiltonian of the wire at equilibrium Eq.(\ref{Hdecomposed}), they depend on both temperature and other length scales such as $\omega_L$ (Eq.(\ref{omegaL})).
 The important remark we have to make is that in the oscillating factors, $\omega(x)$ appears only in front of the difference $t_1-t_2$, while $\omega_{12}$ is in front of all other times.

In the limit $\omega(x)\gg T$, rapid oscillations with respect to $t_1-t_2$ allows to retain only  $|t_1-t_2|\ll 1/T$ as in the pure wire case recalled above. This amounts to take the zero temperature limit in the function $g(t_1-t_2)$ in Eq.(\ref{formalexp}). Similarly if one takes $\omega_{12}\gg T$, zero-temperature limit can be undertaken in the function $\chi$ of the other pairs $t_l-t_k$. This has the consequence that a "truly" non-equilibrium situation where the zero-temperature limit can be undertaken corresponds to both:
\begin{equation}\label{out}
\hbar\omega(x),\hbar\omega_{12}\gg k_BT
\end{equation}
for all $x$ where the tunneling amplitude does not vanish.

 But if either $\omega(x)\gg T\gg \omega_{12}$ (respectively $\omega_{12}\gg T\gg \omega(x)$), we have a "mixed" situation. The temperature cannot be discarded: one has simultaneous dependence on $\omega(x)$ (rep. $\omega_{12}$) and temperature of the currents!

  Through this analysis, we have confirmed simultaneously the important criteria for the validity of such expansion, which is the same as in the two-terminal geometry, i. e. Eq.(\ref{criteria})
  where $\omega^*$ is given by Eq.(\ref{omega*}).

  What happens now if we cannot make such an expansion, thus when instead $\omega^*< \omega^*_{nq}$? Even if $\omega(x)> \omega^*_{nq}$, we can draw the crucial conclusion that $\omega(x)$ cannot offer a natural cutoff to stop the RG flow. Thus the wire is driven into the strong scattering regime, and more precisely to its fixed point if it has one! The fact that $\omega(x)>\omega^*_{nq}$ cannot validate a perturbative expansion with respect to $\mathcal{H}_{nq}$, nor even an RG analysis around small $\mathcal{H}_{nq}$! Rather, one could make such expansion around the possible fixed point, in the strong coupling regime. The same qualitative analysis concerning the order of limits is expected to hold.

   Let us comment briefly the role of the finite size. This introduces an energy cutoff for the RG grow if $\omega^*=\omega_L> \omega_{nq}^*$, which allows to do the series expansion. But it does not allow to take the zero-temperature limit in this expansion when one has a grounded wire, or more precisely when one has  low voltage difference across the wire, $\omega_{12}\ll T\ll \omega_L$. Our argument still holds: $\omega_L$ does not offer oscillating factors with respect to time which replace those at $\omega_{12}$. This can be checked in the step-like inhomogeneous model, where the local correlation functions $\chi$ contain a part similar to an infinite wire, where the length does not intervene. In this part, temperature has to be kept finite as well, thus this questions the recent work in \cite{dolcini_stm} in the limit $V_1=V_2$. There terms to order $\Gamma^2\lambda$ where $\lambda$ is the backscattering amplitudes were computed.

   If we consider again this frequently used limit, corresponding to the majority of the experimental setups, letting $\omega_{12}=0$, one sees why special caution is required. As one has always infinitesimal temperatures, this corresponds to have $\omega_{12}\ll T$, thus the temperature dependence cannot be ignored as explained before. This is contrary to theoretical works dealing with perturbation with respect to $\mathcal{H}_{nq}$ where  the zero-temperature limit is taken in the many point correlation functions similar to $\chi$ in Eq.(\ref{formalexp}). If the wire is infinite, power laws with respect to the unique left energy scale $V$ can be induced from a change of variable scaling in the multiple time integrals, $t_l\rightarrow Vt_l$. But the integrals are not guaranteed to converge, and even when they do for a certain range of parameters, the perturbation series does not converge, thus such power law behavior is meaningless. This is because $V$ does not cut the divergence. If no other energy scales play this role, the wire is driven to the strong coupling regime of $\mathcal{H}_{nq}$. Those facts will be illustrated in more details for local backscattering in section \ref{scaling}, where in addition generalized scaling laws are shown to any order with respect to a local backscattering.

   \subsection{Cross-correlations for weak tunneling and for any $\mathcal{H}_{nq}$ }
 \label{subcross}
 We discuss here a crucial consequence of our previous analysis, which has been overlooked in the literature, and which is relevant to more general three-terminal geometries. This is the case especially when they are used to investigate purely shot-noise or quantum statistics of carriers through current-correlations between two terminals, while the third one plays the role of an injector. On one hand, the simplest situation corresponds to applying a voltage $V_3$ to the injector, while the two collectors are at the same voltage, $V_1=V_2$. On the other hand, in order to get rid of thermal noise contributions which mask the signature of quantum statistics, one usually needs a non-equilibrium regime, where temperature is very low with respect to the bias, the latter being controlled by $V_3$.

 Our previous discussion has shown that this is not a truly non-equilibrium situation, because one needs as well to have $\omega_{12}=eg_L(V_1-V_2)/\hbar\gg T$! As the temperature is never zero in realistic situations, $\omega_{12}$ has a lower bound. Still the order $\omega_{12}\gg T$ has to be kept even when one needs to take the limit of very small $T$ and $\omega_{12}$. This means one has to let $T\rightarrow 0$ first then $\omega_{12}\rightarrow 0$. But in our geometry, we have seen that if $\omega^*=max(\omega_{12},\omega_L,T)$ is too small, the wire is driven to the strong coupling regime. Thus in this limit no perturbation can be performed with respect to $\mathcal{H}_{nq}$, even when $\omega(x)$ at all tunneling points $x$ are high enough.

If one takes the opposite order of limits, i. e. letting $\omega_{12}\rightarrow 0$ first then $T\rightarrow 0$, thus $\omega_{12}\ll T$, corresponds to an equilibrium regime in the wire even when $\omega(x)\gg T$.
 Two main problems arise in this situation.

  Firstly, the expressions in Eqs.(\ref{S12},\ref{snqdecomposed}) were indeed obtained when requiring $\omega_{12},\omega(x)\gg T$ in order to get rid of a contribution which depends on both voltages and temperature. If we focus on weak tunneling, with an arbitrary $\mathcal{H}_{nq}$, such contribution contains the opposite of the first two terms in the exact two terminal noise in Eq.(\ref{S12T}). Those terms vanish only for $\omega_{12}\gg T$, see Eq.(\ref{S12snq}).
  One could circumvent those terms by looking at a kind of excess current-correlations given by
  \begin{equation}\label{excess}
  \Delta S(\omega(x),\omega_{12})=S_{12}(\omega(x),\omega_{12})-S_{12}(\omega(x),\omega_{12}=0).
  \end{equation}
 Notice that this could be an interesting procedure to get rid of background undesirable noise. It has not its analog in the zero-frequency two-terminal noise, but only at finite frequency.
 However this would not solve the second crucial problem. As we saw before, the injector is coupled to a wire at equilibrium, thus thermal effects cannot be discarded form the noise expressions. This will be explicitly shown for a backscattering Hamiltonian.

 This has the consequence that if one needs to deal with relevance of the sign of current correlations in "an infinite wire", such as the limit taken in \cite{crepieux_stm}, one requires to keep the voltage difference $\omega_{12}$ much greater than $T$, more precisely to ensure the out-of-equilibrium condition Eq.(\ref{out}). We expect the same important conclusion in the strong coupling regime, at $\omega^*<\omega^*_{nq}$.

 Nevertheless, one could argue that as $\mathcal{H}_{nq}=0$ in Ref.\cite{crepieux_stm}, one has not to worry about the dependence of $S_{12}$ on temperature. This is provided the "excess" correlations in Eq.(\ref{excess}) are undertaken in order to discard the thermal noise of a pure wire, given by $-4k_Bg_LT$. Nevertheless, we will see that there is an effective generated $\mathcal{H}_{nq}^T$ due to virtual opposite tunneling paths (Eq.(\ref{HnqT})) where only $\omega_{12}$ enters. Thus we can conclude that we need again Eq.(\ref{out}) even when looking at a pure wire. This allows to address the quantum statistical signature in the current correlations, as well as to take into account the higher-order tunneling relevant processes in Eq.(\ref{HnqT}).

 When the out-of-equilibrium condition is satisfied, Eq.(\ref{out}), we can draw a general and crucial result valid for arbitrary $\mathcal{H}_{nq}$, and based only on the requirement of weak tunneling amplitudes. As the renormalized tunneling amplitudes are irrelevant in the RG sense, one can ignore them compared to the relevant processes in $\mathcal{H}_{nq}$, thus $S_{12}$ in Eq.(\ref{S12}) is dominated by its exact out-of-equilibrium value in a two-terminal situation, Eq.(\ref{S12two}):
\begin{equation}\label{S12snq}
S_{12}\simeq -s_{nq,nq}.
\end{equation}
This is a non-perturbative result with respect to $\mathcal{H}_{nq}$. For local scattering processes, $s_{nq,nq}$ is the correlator of the same operator at the same point, and can be shown to be positive by a spectral decomposition. In this case one has:
\begin{equation}
S_{12}(\hbar\omega_{12},\hbar\omega(x)\gg k_BT) <0.
\end{equation}

 This is plausible as electrons Tunnel from the injector in the wire, even though the latter is interacting.
\label{cross}

  Throughout the present section, we have considered second-order expansions with respect to tunneling. But there are short-time contributions which were not written. They are independent on the voltage of injection, and resembles the two-terminal current and noise, $I_{nq}$ and $s_{nq,nq}$, in the presence of backscattering. They correspond to the generation of backscattering by tunneling, and can therefore be implicitly incorporated in $\mathcal{H}_{nq}$ as we show next.
\section{Backscattering generated by Tunneling}
\label{sectunnel}
Usually, for an injector without interactions, tunneling process is known to be irrelevant. Here we will show that the tunneling Hamiltonian in Eq.(\ref{HT}) can generate an effective Hamiltonian which could be even more relevant than $\mathcal{H}_T$ itself, and thus induces invasive effects on the wire. For weak tunneling this effect can become important only in the low-energy limit. This questions the perturbative expansion with respect to tunneling. But we will see that it is valid provided the generated backscattering process is incorporated in $\mathcal{H}_{nq}$.

One can see already this effect in the noninteracting limit, where such a three-terminal structure can be described by a $3X3$ scattering matrix, whose unitarity imposes to have a reflection coefficient for electrons coming from terminal $2$, $r_{22}\simeq t_{32} t_{31}^{\dagger}$, where $t_{3l}$ are the transmission coefficients from the injector $3$ to $l=1,2$. They become respectively proportional to the tunneling amplitude from $3$ to $1$ or $2$ given by $\Gamma_-$ or $\Gamma_+$ when the latter are weak.

We can now notice the following: either in Eq.(\ref{Gferminj}) for $r_2=-r_1$, or in Eq.(\ref{deltaInqT}), the short-time difference contribution is of the form $<\Psi_+(x_1,t_1)\Psi_-^{\dagger}(x_2,t_2)>$, thus corresponds to spatially non-local backscattering processes. Besides, as $\omega_+-\omega_-=2\omega_{12}$, see Eq.(\ref{Vr}), one recovers the expected voltage difference.

 One can see this in another way, restricted however to a non-interacting injector, which could have been as well a starting point to express currents. It consists in integrating out the field of the injector $\Psi_{inj}$ in order to get an effective Keldysh action for the wire. Again, we give here the general form without specifying the Keldysh's matrix labels:
  \begin{equation}
S_{eff}=\int\sum_{r_1,r_2} \lambda_{r_1,r_2}(1,2)\Psi_{r_1}^{\dagger}(1) \Psi_{r_{2}}(2)\label{effectivestunnel},\end{equation}
 where $l$ abbreviates $(x_l,t_l)$ and integration over all coordinates is implicit. $\lambda_{r_1,r_2}(1,2)$ was introduced in Eqs.(\ref{lambdar1r2},\ref{Gferminj}). Here again, for short time difference $|t_1-t_2|\simeq 1/\Lambda$ and for $r_1=-r_2$, spatially non-local backscattering terms as in Eq.(\ref{HB}) are generated. Their amplitude $\lambda_{r_1,r_2}(1,2)$ at $t_1\simeq t_2$ is  proportional to
 $\Gamma_+\Gamma^{\dagger}_-$. As one does not disregard $\mathcal{H}_T$ by such integration, one can keep only such short-time contributions for $r_2=-r_1$ as an effective Hamiltonian. Notice that the Tunneling Hamiltonian itself is an effective one! More precisely, in the previous formalism, one has to replace the non-quadratic Hamiltonian in Eq.(\ref{Htot}) by:
 \begin{equation}\label{HnqT}
 {\mathcal{H}_{nq}}\rightarrow {\mathcal{H}_{nq}}={\mathcal{H}_{nq}}+\int dx_1 dx_2 \lambda_{+,-}(x_1,x_2;t_1\simeq t_2)\Psi_+(x_1)\Psi_-^{\dagger}(x_2)+h. c.
 \end{equation}
 Notice that because $t_1\simeq t_2$, and $\omega_+-\omega_-=\omega_{12}$, one gets in $\lambda(1,2)$ the same translation of $\Phi$ by $\omega_{12}t$ as in Eq.(\ref{translation})! Such generation can be viewed as the virtual process at close times of a right-going electron tunneling from a point $x_1$ of the wire into the injector while a left-going one is injected at $x_2$.

  Nevertheless, the integration leading to Eq.(\ref{effectivestunnel}) was implicitly based on a quadratic Gaussian action with respect to the  fermion fields in the injector, thus without interactions in the injector region coupled by tunneling to the wire. But even when the injector is itself another interacting system \cite{takis_stm}, such generation process can as well be shown from an RG analysis, where the backscattering amplitude flow contains a term in $\Gamma_+\Gamma_-^{\dagger}$.
  For this, even though it is not necessary, let us assume for simplicity the wire is modeled by an homogeneous TLL with parameter $g$ and that tunneling is local. Then one can
derive the RG equations coupling the backscattering amplitude $\lambda$ to $\Gamma$ :
\begin{eqnarray}\label{renormalization}
\frac{d\lambda}{dl}&=&(1-g)\lambda+c\Gamma_+\Gamma_-^{\dagger}\nonumber\\
\frac{d\Gamma_{\pm}}{dl}&=&-\frac{\alpha}2\Gamma_{\pm},
\end{eqnarray}
where $c$ is a non-universal function of
$g$. The exponent $\alpha$ is given by Eq.(\ref{alphabulk}) if the injector is noninteracting. $\alpha$ has different values if we allow for interactions in the injector, depending on whether it is coupled  in its bulk or at its end to our wire. Here the cutoff is $\Lambda e^{-l}$. For a finite wire
connected to reservoirs, Eq.\ref{renormalization} gets more complicated but holds roughly for points in the bulk as
long as $\Lambda e^{-l}>v/L$, otherwise one has to replace $g$ by
its value $1$ in the leads\cite{ines_ann,ines_prb_long}. One can
easily integrate this RG system,
\begin{eqnarray}\label{solution}
\lambda(l)&=&\left[\lambda(0)+2c\Gamma_+(0)\Gamma_-^{\dagger}(0)\right]e^{(1-g)l}+c\Gamma_+(l)\Gamma_-^{\dagger}(l).\nonumber\\
\Gamma_{\pm}(l)&=&\Gamma_{\pm}(0)e^{-\frac{\alpha}2 l}.
\end{eqnarray}
Notice however that the way the RG is cut in the out-of equilibrium situation is not fully established. Thus even though irrelevant to lowest order,
tunneling generates relevant backscattering. A similar RG generation occurs in the two-chains problem\cite{yakovenko} or crossed carbon nanotubes\cite{egger_cross}. As tunneling is weak, we expect this generation to be more important at low energy, such that the generated backscattering grows enough. These RG equations can be generalized to extended tunneling and an arbitrary Hamiltonian $\mathcal{H}_{W}$. Thus one has as well to incorporate the nonlocal backscattering processes in $\mathcal{H}_{nq}$ Eq.(\ref{HnqT}). This affects the transport properties even when starting from a pure wire described by $\mathcal{H}_{q}$ only.

Notice that there is no such generation for uni-directional injection such as that performed in \cite{yacoby_fraction}, thus this procedure would be the safest way to probe the bare Green's functions, restricted however to Fermi fields with the same "injected" direction, i. e. $r_1=r_2$ in Eq.(\ref{IrGamma}). This would be sufficient if the wire is pure but also homogeneous, such that $<\Psi_+\Psi_-^{\dagger}>_W=0$. It requires also the injector to be far enough such that Coulomb interaction with the wire considered next can be neglected.


\section{Coulomb interactions between the wire, the injector and the gate}

\label{invasive}
This section is intended to show qualitatively how invasive effects in the wire arise by its Coulomb interactions with the injector.  We will not specify the form of such interactions, neither the shape and the extension of the injector and gate. The present analysis can be applied either to sharp or extended STM.
We give rather general qualitative conclusions and starting formalism to explore realistic situations. Usually it is believed that the Coulomb interactions can be ignored when the distance between the injector and the wire is much higher than the Fermi wavelength. Nevertheless, the present approach might offer a potential way to check the validity of this semi-classical argument. Screening by a gate can be treated in a similar way. Nevertheless, if they have a regular extension, they shouldn't introduce scattering processes but merely screening.

We start by the injector, the same analysis will be reproduced briefly for the gates.
The underlying idea is to integrate out the injector, in the spirit of \cite{ines_saleur} for an ohmic environment. Strictly speaking, if different voltages are applied to the injector and the two reservoirs to which the wire is connected, one needs a Keldysh action. Nevertheless, to simplify the presentation, we will restrict to an Euclidian action which gives the same type of terms. We also assume that the injector density modes are plasmonic, being described by a Gaussian action, and that they commute with the Fermi field that enters in tunneling. This is plausible if charge relaxation after a tunnel event is rapid enough, and is similar to approximations made when the coupling to an environment is treated.

However, this restricts the generalization to an injector formed by another one-dimensional interacting systems, unless one ensures some conditions. In this case, the density in the injector is as well given by Eq.(\ref{rhotot}). It contains the plasmonic part $\rho$ but other short-wavelength components as well. Nevertheless, the coupling to $\rho$ to which we restrict ourselves was not included in previous works on Coulomb drag or crossed nanotubes, to which our analysis might be relevant.

We will show how Coulomb interactions with the injector can induce both inhomogeneous screening and non-local backscattering processes, adding to those already generated by tunneling in Eq.(\ref{HnqT}). Thus all terms in Eq.(\ref{Htot}) get corrected: the injector can renormalize the parameters $v(x)$ and $g(x)$ of $\mathcal{H}_{q}$ in Eq.(\ref{H0}), generates possible finite-range interactions $U(x,y)$, as well as corrections to $\mathcal{H}_{nq}$ which add to those due to tunneling in Eq.(\ref{HnqT}).

We have written the Hamiltonian term for the Coulomb interactions between the wire (labeled by W) and the injector in Eq.(\ref{Winj0}) which adds to Eq.(\ref{Htot}):
  \begin{equation}\label{Winj}
   \mathcal{H}_{W,inj} = \int_{W} dx \int_{inj}d{{\vec{r}}}
U_{Coul}(x,{\vec{r}}) \rho_{tot}(x)\rho_{inj}({\vec{r}}),
\end{equation}
where $\rho_{tot}(x)$ is given by Eq.(\ref{rhotot}) and $\rho_{inj}$ is the density in the injector we write again here: \begin{equation}\label{rhoinj}\rho_{inj}(\vec{r})= \left<\rho_{inj}\right>+\delta\rho_{inj}(\vec{r}),\end{equation}containing, besides the density fluctuations $\delta\rho_{inj}$, the average value of the excess density. We have seen how both terms modify $V(x)$ in Eq.(\ref{V(x)}), and we will consider the remaining effects here.
 Let us consider again the role of the average density:
\begin{equation}<\rho_{inj}>\simeq e v_F V_3.\end{equation}
Its contribution to Eq.(\ref{Winj}) reads:
$
\int V_{inj}(x)\rho_{tot}(x),
$
where
\begin{equation}\label{Vinj}V_{inj}(x)=ev_{F,inj} V_3 \int_{inj}
U_{Coul}(x,{\vec{r}})d{{\vec{r}}}.
\end{equation}
Using Eq.(\ref{rhotot}), one gets a coupling term to the long wavelength part $\rho(x)$ which was taken into account in $V(x)$ inside the wire, Eq.(\ref{V(x)}). Most importantly, one gets also a backscattering term, similar to Eq.(\ref{HB}) with $\lambda(x)$ given rather by $V_{inj}(x)$, Eq.(\ref{Vinj}). Thus if the bare backscattering was already present in $\mathcal{H}_{nq}$, i. e. if the latter contains Eq.(\ref{HB}), one has to replace: \begin{equation}\label{lambdainj}\lambda(x)\rightarrow \lambda(x)+V_{inj}(x).
\end{equation}
Let us now consider the contribution of the injector density fluctuations in Eq.(\ref{rhoinj}) to $\mathcal{H}_{W,inj}$, Eq.(\ref{Winj}), which we will integrate out. For this, we assume $\delta\rho_{inj}$ to be described by a gaussian action:
$$S_{plas}=\frac 1 2\int_{inj}\int \delta\rho_{inj}(\vec{r},\tau)G_{plas}^{-1}(\vec{r},\vec{r}',\tau-\tau')\delta\rho_{inj}(\vec{r}',\tau'),$$ where $\int$ implicitly denotes integration on double space and imaginary time coordinates. Integration yields an effective action term for the wire:
\begin{equation}\label{heff}
 \delta S_{eff}=
\frac 1 2\int_W \int \rho_{tot}(1)U_{eff}(1,2)\rho_{tot}(2),
\end{equation}
where $l$ refers to $(x_l,t_l)$ and
\begin{equation}\label{Ueff}
U_{eff}(1,2)=U_{eff}(x_1,x_2,\tau_1-\tau_2)=\int\int d\vec{r}_1d\vec{r}_2U_{Coul}(x_1,\vec{r}_1)G_{plas}(\vec{r}_1,\vec{r}_2,\tau_1-\tau_2)U_{Coul}(x_2,\vec{r}_2).
\end{equation}
 Indeed, at equal times, Eq.(\ref{heff}) has a similar form to the most general interaction Hamiltonian inside the wire. Under some conditions, these reduce to coupling between the long-wavelength parts of the density components, as in Eq.(\ref{H0}). Nevertheless, this is no not necessarily the case for general inhomogeneous $U_{eff}$. A detailed analysis was performed in \cite{ines_ann}, and is exploited here in order to quote the different kinds of terms one gets by implementing the expression of the total density in Eq.(\ref{rhotot}).
Performing integrations by part, one can write Eq.(\ref{heff}) as:
\begin{eqnarray}\label{explicit}
\delta  S_{eff}&=&\frac{1}{\pi^2}\int U_{eff}(1,2)
\partial_{x_1}\Phi(1)\partial_{x_2}\Phi(2)\nonumber\\&&-\partial_{x_1}\partial_{x_2} U_{eff}(1,2)\sum_{m_1,m_2\neq 0}\frac 1{4m_1 m_2}e^{2i\sum_{l=1,2}m_l[k_F x_l+\Phi(l)]},
 \end{eqnarray}
where again $l$ denotes $(x_l,t_l)$. We see that the first term yields a local renormalization of the bare quadratic and nonlocal spatial interactions $U$ in the associated action to $\mathcal{H}_q$ in Eq.(\ref{H0}), but with additional non-locality in imaginary time. This is the case as well for the terms with $m_1+m_2=0$ which can be approached by two terms: a linear and a quadratic form in $\Phi$. The terms with $m_1+m_2\neq 0$ correspond to non-conserving momentum processes which could be neglected only if $U_{eff}$ varies slowly in space on the scale of $\lambda_F$, due to the oscillating factors $e^{2ik_Fmx}$.

 Now let us consider the domain where $|\tau_1-\tau_2|$ is very small, less than $1/\Lambda$. As one can approach $\tau_1\simeq \tau_2$, one is left with  a unique integration over $\tau$, and the corresponding generated terms in the effective action (\ref{explicit}) can be reduced to a correction to the total Hamiltonian. Indeed, such domain is probably the most dominant in $S_{eff}$ in the RG sense whenever the plasmonic Green's function thus $U_{eff}$ decay fast enough in time, see Eq.(\ref{Ueff}). In this case, the additional effective Hamiltonian has the same form as Eq.(\ref{explicit}) where imaginary time integrations are dropped, while $U_{eff}$ is taken at $|\tau_1-\tau_2|\simeq 1/\Lambda$. This yields four types of corrections:
 \begin{enumerate}
 \item
  Due to the first term on the r. h. s. of Eq.(\ref{explicit}), $U_{eff}$ adds to $U$ in Eq.(\ref{H0}):
 \begin{equation}\label{Uren}
 U(x_1,x_2)\rightarrow U(x_1,x_2)+U_{eff}(x_1,x_2,1/\Lambda),
 \end{equation}
 which can be viewed as a screening effect, see Eq.(\ref{Ueff}).
 \item
  In the second term on the r. h. s. of Eq.(\ref{explicit}), and for $m_1+m_2=0$, one can make an expansion in $\partial_x\Phi$. The linear term yields an inhomogeneous potential $\delta_{inj} V(x)$ which couples to $\rho(x)$. It was included in $V(x)$, Eq.(\ref{V(x)}). It is analogous to that generated by similar internal interactions in the wire \cite{ines_ann} in Eq.(\ref{deltaV(x)}):

  \begin{equation}\label{deltaV(x)inj}
  \delta V_{inj}(x)=\int \frac{dy k_F^2}{\pi} y\bar{U}_{eff}(x,y,1/\Lambda) \sin 2k_Fy,
  \end{equation}
  where the bar indicates the change of variables in Eq.(\ref{barF}).
\item The next term in the contribution of $m_1+m_2=0$ is in $(\partial_x\Phi)^2$. It adds inhomogeneous corrections to $v(x)/g(x)$ in Eq.(\ref{H0}):

\item
Finally terms with $m_1+m_2\neq 0$ on the r. h. s. of Eq.(\ref{explicit})
yield spatially nonlocal backscattering processes, thus gives contribution to the non-quadratic Hamiltonian similar to Eq.(\ref{HnqT}):

\begin{equation}\label{HnqCoul}
\mathcal{H}_{nq}\rightarrow \mathcal{H}_{nq}+\int dx_1dx_2
\partial_{x_1}\partial_{x_2} U_{eff}(x_1,x_2,\lambda_F/v_F)\sum_{m_1+ m_2\neq 0}\frac 1{4m_1 m_2\pi^2}e^{2i\sum_{l=1,2}m_l[k_F x_l +\Phi(x_l)]},
\end{equation}
   where $U_{eff}$ is given in Eq.(\ref{Ueff}).

\end{enumerate}
Recall that the injector can be either a sharp or an extended tip. As we have noticed, in case it is a TLL, even though other terms might be generated, the connection of our analysis to works on Coulomb drag or crossed nanotubes would be interesting to do \cite{nazarov_drag,flensberg}. We have to notice that in the domain where backscattering terms or such corrections to $\mathcal{H}_{nq}$ are spatially extended, and are treated by an RG approach, they can as well renormalize the interactions in Eq.(\ref{H0}).

A similar treatment of the gates can be performed as well. The average excess density in the gate is included in $V(x)$ (see Eqs.(\ref{V(x)gate},\ref{V(x)two})). The fluctuating part can be integrated out as above. One expects nevertheless the backscattering type terms to be much weaker, unless there are for instance sharply ended gates. One mainly gets screening terms similar to Eq.(\ref{Uren}), where $U_{eff}$ is given by Eq.(\ref{Ueff}) with integration over the gate, and where the plasmonic Green's function is taken in the gate as well.
 \section{Case of a sharp injector}

 \label{sharp}
  Let us now consider the special situation where the injector is a sharp tip of extension of order of few wavelengths, coupled to a point $x_0$ in the lower wire, as in Fig.(\ref{sharp}). This restricts the spatial coordinates $x_1,x_2$ to a small region around $x_0$ and the corresponding end point of the tip for instance in: the Tunneling Hamiltonian in Eq.(\ref{HT}), the Coulomb interactions in Eq.(\ref{Winj}), the corrections due to tunneling in Eq.(\ref{HnqT}) and those due to Coulomb interactions in Eqs.(\ref{HnqCoul},\ref{Uren}). Sharpness makes less spatially extended the inhomogeneous screening, in Eq.(\ref{Uren}) as well as the effect on $v(x)/g(x)$ mentioned above. At the same time it increases the $2k_F$ components of the different functions intervening in backscattering type terms. This can be especially important in case the tip is close to the  wire. They resemble a local impurity amplitude if one ignores the non-locality in time, assuming the plasmonic Green's function in the injector to decay fast enough with time. Thus there are on one side local changes in interactions in $\mathcal{H}_q$, Eq.(\ref{H0}). On the other side there are corrections to the non-quadratic bare Hamiltonian $\mathcal{H}_{nq}$ in Eq.(\ref{Hdecomposed}) by a local backscattering Hamiltonian, $\mathcal{H}_{B}$, in Eq.(\ref{HBlocal}). The amplitude $\lambda$ is the combination of three types of backscattering processes at $x_0$:
\begin{enumerate}
 \item
That with an amplitude proportional to $\Gamma_+\Gamma_-^{\dagger}$, in Eq.(\ref{HnqT}), more precisely from the local $\lambda_{+,-}$ in Eq.(\ref{lambdar1r2}):

 \begin{equation}\label{lambdatunnel}
 \lambda_{tunnel}=\int\int dx dy\lambda_{+,-}(x,y,0)e^{i [k_Fx+k_Fy+\phi_0(x)+\phi_0(y)-\theta_0(x)+\theta_0(y)]}.
 \end{equation}
\item That in Eq.(\ref{lambdainj}) where the amplitude is given by $V_{inj}$ in Eq.(\ref{Vinj}):
\begin{equation}
\lambda_{inj}=\int V_{inj}(x)e^{2i(k_Fx+\phi_0(x))}
\end{equation}
\item the Coulomb interactions with the injector written in Eq.(\ref{HnqCoul}), whose contribution to $\lambda$ reads, focusing again on the most dominant backscattering term:
\begin{equation}
\lambda_{coul}=\sum_{m_1+ m_2=\pm}\int dx_1dx_2
\partial_{x_1}\partial_{x_2} U_{eff}(x_1,x_2,\lambda_F/v_F)\frac 1{4m_1 m_2\pi^2}e^{2i\sum_{l=1,2}m_lk_F x_l+\phi_0(x_l) },
\end{equation}
  where integration runs over the small space domain around $x_0$.
  \end{enumerate}
  Thus:
  $$\lambda=\lambda_{tunnel}+\lambda_{inj}+ \lambda_{coul}.$$
  We have expressed (section \ref{weaktunneling}) the currents to lowest order in tunneling, without expansion in $\mathcal{H}_{nq}$ which contains now $\mathcal{H}_{B}$. We have also shown that  $\omega(x_0)=\omega_+(x_0)+\omega_-(x_0)$ (see Eqs.(\ref{HT},\ref{omega(x)},\ref{Vr})) does not prevent the growing of relevant processes. In particular a perturbative expansion with respect to $\mathcal{H}_{B}$ requires either high enough T or $\omega_L=v/L$ if one lets $\omega_{12}=0$.
 \subsection{What does an STM probe coupled to a wire measure?}
 Let us now take our injector as an STM. As both virtual second-order tunnel processes and possible Coulomb interactions generate a local backscattering Hamiltonian $\mathcal{H}_{B}$, and the density Hamiltonian in $\mathcal{H}_{q}$ is as well modified on local scales around $x_0$,  the STM affects itself the DOS it measures! It is no more the expected  bulk density of a pure wire, but that in the presence of an effective impurity, with possible change in the exponents as well.

 The formal expressions of the current we derived can be used. They don't rely necessarily on perturbation with respect to backscattering. Now $j_{nq}$ in Eq.(\ref{jnq}) vanishes apart from $x=x_0$ and is simply given by:
 \begin{equation}
  j_{nq}=k_F\lambda e^{2i(\omega_{12}t+i\Phi(x_0))}+h.c.
  \end{equation}

For a DOS measurement, tunneling has to be weak enough, thus the currents at the contacts in Eq.(\ref{I12}) can be expanded to lowest order using the expressions of $I_r$ and $I_{nq}^T$ in Eqs.(\ref{IrGamma},\ref{Idecomposed},\ref{deltaInqT}). Recall that such second-order perturbation with respect to tunneling is valid because the generated backscattering amplitude in Eq.(\ref{lambdatunnel}) is now incorporated in the effective $\mathcal{H}_{B}$. In particular,
 the average of $j_{nq}$ in the two-terminal geometry, yielding $I_{nq}$ on the r. h. s. of Eq.(\ref{Idecomposed}), can be obtained from the two-terminal differential conductance considered in the first section. For weak enough $\lambda$, this can be expanded perturbatively with respect to $\lambda$ in two limits through Eqs.(\ref{coherent},\ref{incoherent}).
  As we have already shown  for a general non-quadratic Hamiltonian $\mathcal{H}_{nq}$ (see section \ref{weaktunneling}), the validity criteria for such perturbation is the same as that of the two-terminal geometry, it implies the same $\omega^*$ in Eq.(\ref{omega*}): $\omega^*>\omega_B^*$, where $\omega_B^*$ is the energy crossover to the strong-backscattering regime. It is given for instance by Eq.(\ref{omegaB*}) if the local screening is ignored and interactions are uniform and short-range. Taking equal voltages on both reservoirs as often done in a DOS measurement, thus $\omega_{12}=0$, requires to keep either $T$ or $\omega_L$  $>\omega_B^*$ : it is not sufficient in this case to have the voltage $\omega(x_0)$ higher than $\omega^*_B$ to stop the backscattering RG flow as believed in related literature. If not, thus if $\omega^*=\max(\omega_{12},k_BT/\hbar,\omega_L)< \omega_B^*$, letting $\omega(x_0)> \omega_{B}^*$ does not prevent the wire from going to its low energy fixed point. Here it corresponds to being disconnected at $x_0$, see Fig.(\ref{figcoupe}). Therefore, we draw the crucial following conclusion: it is the end DOS and not the bulk expected one, that the STM probes.

 What happens if additional backscattering processes are present, for instance the contacts are not perfectly ohmic as is often the case? There is another crossover energy $\omega_B^{*'}$ below which one reaches the strong backscattering regime, thus where tunneling contacts are to be considered. They are controlled by the end exponent to a noninteracting lead, given by Eq.(\ref{alphaend}) for the case of a homogeneous TLL. At $\omega^*$ smaller than both $\omega^*_B,\omega_B^{*'}$, the system is disconnected into two segments. In case the injector is noninteracting, it probes the same end exponent as that at the contacts.

 We have also shown that for the cross-correlations to not contain thermal noise contribution, thus can be related to quantum statistics, we need both $\omega_{12},\omega(x_0)\gg T$. In this case, as in Eq.(\ref{S12snq}), they are dominated by the opposite of the backscattering noise correlations of $j_{nq}$ in the presence of $\mathcal{H}_q+\mathcal{H}_B$ only:
 $$S_{12}\simeq -s_{nq,nq}.$$
 This relies only on the weak tunneling amplitude. We don't need to assume $\mathcal{H}_B$ to be weak. Thus this is the exact non-equilibrium noise in a two-terminal geometry at $T\ll \omega_{12}$, and is valid in both the weak as well as the strong backscattering regime. If one starts from a pure wire, or just local  $\mathcal{H}_{nq}$ at the tunneling point $x_0$ to which one adds the generated backscattering process, $s_{nq,nq}$ is the correlator of the same current operator at the same position. A spectral decomposition allows to show that it is positive. Therefore:
 \begin{equation}
 S_{12}<0.
 \end{equation}
 Notice that local coupling to the injector corresponds to the most relevant situation to study signatures of quantum statistics. Otherwise, with an extended tunneling, there are complicated spatial interference effects. In particular, this solves the paradox of Ref.(\cite{crepieux_stm}): in the regime the authors consider of a pure LL with parameter $g$ and where Coulomb interactions with the injector are not included, the generated amplitude is of order $\Gamma_+\Gamma_-^{\dagger}$. Thus in the weak backscattering regime one has:
 $$
 S_{12}\simeq -c^2|\Gamma_+|^2|\Gamma_-|^2\left|\frac{\omega_{12}}{\Lambda}\right|^{2g-1},$$
 where $c$ intervenes in the RG equations (\ref{renormalization}). In the strong backscattering regime illustrated in Fig.(\ref{figcoupe}), one has to lowest order in the tunneling amplitudes a similar result to the scattering approach, with tunneling amplitudes renormalized by interactions. In particular, it is again negative.
 Notice that one could as well consider the limit of weak interactions which renormalizes the $3X3$ scattering matrix. Again, one can express $S_{12}$ through the scattering theory, to recover a negative sign as in the non-interacting case.

  \section{Novel scaling laws}
   \label{scaling}
   This section illustrates in a more concrete way some of the previous crucial results. Even if it is specific to the sharp electrode in \ref{sharp}, we present it in a separate section as it is important to illustrate the conclusions obtained in \ref{weaktunneling}. The results of the present section applies more generally to an injector with interactions coupled either at its end or in its bulk to the 1-D system. They can be extended to three edge states in the FQHE, such as those in Refs.\cite{heiblum_dilute,kane_fisher_dilute}.
   We adopt here both local tunneling and backscattering in Eq.(\ref{HBlocal}) at the same point $x_0$. Thus $\mathcal{H}_{nq}=\mathcal{H}_B$, Eq.(\ref{HBlocal}). There are three main reasons for our choice:

   \begin{enumerate}
   \item This gives less cumbersome expressions, as there are no multiple space integrals.
   \item It is relevant for the sharp injector in a pure wire, where invasive effects lead to a local backscattering at the tunneling point as  discussed in section \ref{sharp}.
   \item When letting $\omega_{12}=0$ and the length of the wire infinite, it allows to show clearly deviations from scaling laws. These are usually expected when only two energies are left: the voltage of injection and temperature. Allowing for an additional spatial extension would have introduced additional energy scales.

   \end{enumerate}
  In the latter case, in an infinite homogeneous pure TLL with $\omega_{12}=0$, the scaling law obeyed by
  the differential tunneling current for a local tunneling with $\Gamma_+=\Gamma_-=\Gamma$ is:
\begin{equation}\label{ITbulk}
\frac {dI_T}{dV}\simeq |\Gamma^*(T)|^2 f\left(\frac V T\right),
\end{equation}
where the renormalized tunneling amplitude is given by:

\begin{equation}
 \Gamma^*(\omega)=\Gamma \left(\frac{\omega}{\Lambda}\right)^{\alpha}.
 \end{equation} Here $\alpha$ is given by Eq.(\ref{alphabulk}) if the injector is noninteracting and the wire is a TLL, but has other values for an injector given by an interacting wire coupled either at the level of its end or its bulk. In particular, this yields a power law of the differential tunneling conductance with the same exponent $\alpha$ on voltage or temperature if either $\omega(x_0)\gg T$ or $T\gg \omega(x_0)$. This behavior is similar to that in the two-terminal conductance in the presence of backscattering, in Eq.(\ref{scaling12}), with a different exponent. It is important now to know how those laws are modified in the presence of both tunneling and backscattering.

   Even though no proposal has been made to our knowledge, it is implicitly assumed in the related theoretical and experimental work that one expects a scaling law between the voltage injector and the temperature if only those two energy scales are left when one lets $\omega_{12}=0$. Deviations were however observed for instance in Ref.\cite{bockrath_99} without any reliable explanation. Here, we show that deviations occur in the presence of $\mathcal{H}_{B}$. We consider here the regime of weak backscattering. We give the form of the perturbative series, thus we need that either $\omega_{12}$ or $T$ to be greater than $\omega_B^*$. We expect a similar behavior with different exponents in the strong backscattering regime, where the expansion can be made with respect to a Tunneling Hamiltonian at $x_0$ between the two disconnected wires.

 All correlations of currents to second order in tunneling can be expressed in terms of three types of series. To avoid too many notations, we will not introduce different labels for all different series which come into play, but use the same notation for those with a similar behavior.
 The first type of series correspond simply to the two-terminal situation, as the two-terminal current $I_{nq}$ and noise $s_{nq,nq}$ appeared in Eqs.(\ref{Idecomposed},\ref{snqdecomposed}). They are given by Eqs.(\ref{vois},\ref{scaling12}).

 The second type of series comes from expanding the Green's functions $<\Psi_r\Psi^{\dagger}_{ r}>_W$ between co- propagating electrons for $r=\pm$ in the presence of $\mathcal{H}_W(t)=\mathcal{H}_q+\mathcal{H}_B(\Phi+\phi_0(x_0)+\omega_{12}t)$. It contains even powers with respect to the backscattering amplitude:
    \begin{eqnarray}\label{Arr}
    A_{r,r}&\simeq& |\Gamma_r^*(T)|^2\sum_{n=0}^{\infty}\lambda^{*}(T)^{2n}TA^r_{2n}\left(\frac {\omega(x_0)}T,\frac{\omega_{12}} T\right).
   \end{eqnarray}
  The second series comes from the anti-propagating Green's functions $<\Psi_r\Psi_{-r}>_W$ and has odd powers of backscattering amplitudes:
   \begin{eqnarray}\label{Ar-r}
   A_{r,-r}&\simeq& \Gamma_+^*(T)\Gamma_-(T)^{\dagger*}\sum_{n=0}^{\infty}\lambda^{*}(T)^{2n+1}TA^{r}_{2n+1}\left(\frac {\omega(x_0)}T,\frac{\omega_{12}} T\right).
   \end{eqnarray}
   Here the renormalized backscattering and tunneling amplitudes intervene when cutting their flow at $T$. But the functions of voltage differences can change this, as in Eq.(\ref{scaling12}).

  In the limit $\omega(x_0)\gg T$, one can show that a crucial consequence is that the temperature dependence cannot be dropped, as those two expressions become:

  \begin{eqnarray}
 A_{r,r}(\omega(x_0)\gg T)&\simeq& \omega(x_0)|\Gamma_r^*(\omega(x_0))|^2\sum_{n=0}^{\infty}\lambda^{*}(T)^{2n}a^{r}_{2n}\left(\frac{\omega_{12}} T\right)\nonumber\\
A_{r,-r}(\omega(x_0)\gg T)&\simeq& \omega(x_0)\left(\frac T  {\omega(x_0)}\right)^g\Gamma_+^*(\omega(x_0))\Gamma_-(\omega(x_0))^{*\dagger}\sum_{n=0}^{\infty}\lambda^{*}(T)^{2n+1}a^{r}_{2n+1}\left( \frac{\omega_{12}} {T}\right),
  \end{eqnarray}
 where we had to restrict ourselves, in the second expression, for simplicity, to a uniform TLL with a parameter $g$. In particular, for $\omega_{12}=0$, it is the temperature which cuts the flow of $\lambda$. Those series have dependence on both $\omega(x_0)$ and $T$ even though $\omega(x_0)\gg T$. Let's for instance consider the term in $\lambda$, which in $A_{r,-r}$ corresponds to $n=0$. If we had taken the zero-temperature limit in the corresponding integral, one would have had to replace:
 $$\left(\frac T  {\omega(x_0)}\right)^g\lambda^*(T)\rightarrow \lambda^*(\omega(x_0)).$$
In particular, this questions the limit $\omega_{12}=0$ undertaken in some situations in Ref.\cite{dolcini_stm} for two impurities at the contacts.
This is in accordance with our previous crucial conclusion: that the fully out-of-equilibrium situation correspond to both $\omega(x_0)$ and $\omega_{12}$ $\gg T$. As one could as well make similar expansions with respect to a tunneling Hamiltonian between two semi-infinite wires in the strong-backscattering regime, we expect still those facts to hold.

 Let us now specify how the above types of series intervene in the currents in Eqs.(\ref{I12},\ref{IrGamma},\ref{Idecomposed}) and the noise.

We have already seen that the two-terminal bare contributions $I_{nq}$ in Eq.(\ref{Idecomposed}) and $s_{nq,nq}$ in Eq.(\ref{snqdecomposed}), given by Eqs.(\ref{vois},\ref{scaling12}). Notice one can get rid of $s_{nq,nq}$ by the procedure in Eq.(\ref{excess}). Besides that, one has:
\begin{eqnarray}
I_r &\rightarrow &A_{r,r}+A_{r,-r}\nonumber\\
 \delta I_{nq}^T&\rightarrow & \sum_{r=\pm} A_{r,r}+A_{r,-r}.
\end{eqnarray}
Therefore the currents at the contacts, Eq.(\ref{I12},\ref{Idecomposed},\ref{Ir}) contain the three types of series.
For the noise one has:
\begin{eqnarray}\label{sseries}
s_{rr'} &\rightarrow & A_{r,r'}\nonumber\\
 \delta s_{nq,nq}^T&\rightarrow & \sum_{r=\pm} A_{r,r}+A_{r,-r}\nonumber\\
 s_{r,nq}&\rightarrow& A_{r,r}+A_{r,-r}.
\end{eqnarray}
Thus when $\omega_{12}=0$, a perturbative expansion of noise still contains temperature. This illustrates the features presented in section \ref{weaktunneling} for a general non-quadratic Hamiltonian $\mathcal{H}_{nq}$. The perturbative regime with respect to $\mathcal{H}_{nq}$ where the role of charge and quantum statistics can be investigated, requires both $\omega_{12},\omega(x_0)\gg T$ (Eq.(\ref{out})).
  \section{Some concluding remarks}
  \label{conclusion}
  We refer to section \ref{summary} for the main points shown in this paper. Here we recall the main conclusions and their relevance for other related works.

We have shown the presence of invasive effects due to tunneling and Coulomb interactions between the injector and the wire which screen in addition  the interactions in the wire. The tunneling effects apply as well to the case when the injector is an interacting wire, in particular they yield a unique fully disconnected junction when three  TLL arms are considered, contrary to Ref.(\cite{sen_1}). The effect of Coulomb interactions cannot be systematically generalized to this case, unless one requires some conditions. The contribution we have considered was overlooked in works on Coulomb drag or crossed wires, and could be relevant for those setups as well.
The invasive effects due to tunneling are absent for uni-directional injection \cite{yacoby_fraction}. Nevertheless, such possible probe would allow to access only the Green's functions between co propagating electrons, thus is suited only for a pure wire with homogeneous interactions. Even more, we think that screening effects due to Coulomb interactions between the wire and the injector are of relevance for the experimental setup.

 When one starts from a pure TLL and a sharp STM, the measured DOS can be strongly modified in the low energy sector by these invasive effects. The local screening effects could modify the power exponents even though one expects this effect to be negligible. In the corresponding strong backscattering regime, the wire is disconnected at the tunneling location and the STM probes the end DOS instead of the bulk one. Besides, we showed that the STM voltage cannot offer a cutoff to stop the flow to the strong coupling regime.

This questions the work in \cite{glazman_stm}, where an STM probes a semi-infinite wire with a local backscattering term: there is no finite length neither voltage in the backscattering term to cut its flow, and the temperature was set to zero. It questions as well the limit of grounded wire and zero-temperature limit undertaken in some situations of Ref.\cite{dolcini_stm}.

   We can give an additional proof that the  voltage entering in tunneling does not provide an RG cutoff in the limit of a weakly interacting wire, adapting the approach of Lal {\it et al}\cite{sen_1} to our setup. As this work was performed in the equilibrium limit, when the temperature is greater than all the voltage differences, it is not appropriate to discuss the features we have shown. We had rather to perform an original out-of equilibrium RG for a $3X3$ scattering matrix with a general  parametrization. On one hand, this has allowed us to check the generation of backscattering terms in $\Gamma_+\Gamma_-$ as stated above. On the other hand, we have succeeded to check that when the injector is simulated by a semi-infinite non-interacting wire, its voltage does not offer an energy scale to stop the RG\cite{ines_benoit}. Nevertheless, the validity of a first-order expansion with respect to short-range Coulomb interactions $U$ inside the wire is questioned when one starts from weak backscattering $\lambda$. This is because the bulk DOS exponent in a uniform TLL is :
 \begin{equation}
 2\alpha_{bulk}=\frac 1 g+g-2\simeq O(U^2),
 \end{equation}
 thus the bulk exponent would be captured only by a second order RG. Such remark has to be taken seriously into account for the general case, when an interacting semi-infinite wire is weakly coupled to the bulk of another wire, which corresponds to one of the fixed points of the RG flow for the $3X3$ matrix in \cite{sen_1}. Our analysis showed that it is not, as it is driven to the completely disconnected junction.

 Let us mention that when the finite wire enters explicitly into account, it adds another possible energy to cut the RG, and induces oscillations in I(V) when step-like short range interactions are considered\cite{dolcini_05,dolcini_stm}.

 One of the other important messages which have been overlooked in previous three-terminal non-equilibrium regime is that the zero-temperature limit requires not only to have one voltage difference larger than the temperature, but two. This sheds light on the peculiarities of Ref.(\cite{kane_fisher_dilute}) for three edge states in the FQHE. In our model, those are given by $\omega_{12}=eg_L(V_1-V_2)$ and $\omega(x)$ at each tunneling location $x$ expressed in Eq.(\ref{omega(x)}).

 We have also analyzed the cross-correlations. In order to get rid of thermal fluctuations, one needs a "truly" out-of-equilibrium situation, where the voltages entering in the two tunneling Hamiltonians are greater than temperature. In the present setup, we have solved the paradox of positive cross-correlations obtained in Ref.\cite{crepieux_stm}. We have allowed for a general Hamiltonian with arbitrary  inhomogeneous interactions of any range and profile, as well as relevant non-conserving momentum processes, including screening effects of the injector and generation of backscattering by tunneling. The cross-correlations are dominated by their exact two-terminal value. This is the opposite of the auto-correlations which are positive for local processes. Thus the cross-correlations are negative by current conservation in the two-terminal setup. This result holds even if one starts from a pure wire as in \cite{crepieux_stm} due to the induced invasive local effects .

 We have offered novel modified scaling laws for current correlations to all orders with respect to a local weak backscattering potential, and expect similar ones in the SBS as well. Due to the invasive terms we have shown, this behavior is valid in the case a sharp STM probes a pure TLL. This offers a possible path to explain the low temperature deviations from scaling laws seen in the experimental search for TLL behavior through two tunneling probes\cite{bockrath_99}. These laws can be extended to a junction of different TLLs as well as to three edge states in the FQHE. Thus they apply to the setup for studying noise of dilute quasi particles \cite{heiblum_dilute} and shed light on its theoretical analysis in Ref.\cite{kane_fisher_dilute}. They will be analyzed in more details in the future.

Note: While the present paper was being prepared, appeared two recent papers by M. Guigou et al, cond-mat/0904.4019 and cond-mat/0905.3842 with a detailed analysis of the role of Coulomb interactions between a sharp tip and a uniform short-range interacting wire in the DOS and transport.

The author thanks B. Dou\c cot for fruitful discussions and on-going collaboration on this problem. She also thanks C. Bena, J. N. Fuchs,  F. Piechon and P. Simon for discussions and critical reading of the paper, as well as H. Bouchiat, A. Cr\'epieux, R. Deblock and F. Hekking.


\newpage

\begin{figure}
\vspace{0.3cm}
\begin{center}
\caption{ An injector, which can be either a tip or another wire,
coupled by both asymmetric tunneling amplitudes $\Gamma_{\pm}(x,\vec{r})$ to an interacting one-dimensional wire, and by Coulomb interactions $U_{coul}$ treated in section \ref{invasive}, see Eq.(\ref{Winj}). The wire has inhomogeneous interactions of arbitrary form and range described by the functions $v(x),g(x),U(x,y)$ which intervene in its quadratic Hamiltonian part $\mathcal{H}_q$ in Eq.(\ref{H0}), and non-conserving momentum processes described by the non-quadratic Hamiltonian $\mathcal{H}_{nq}$. It is connected at $\pm L/2$ to leads which are non-interacting in realistic setups, but where we keep short-range interactions with parameter $g_L$. The three voltages intervene through the two main frequencies $\omega_{12}=[\omega_{+}(x)-\omega_-(x)]/2$ and $\omega(x)=\omega_+(x)+\omega_+(x)$, see Eqs.(\ref{omega12},\ref{omega(x)},\ref{Vr},\ref{V(x)}), which will determine the out-of-equilibrium criteria in Eq.(\ref{out}).}
\end{center}
\end{figure}
\begin{figure}
\vspace{0.3cm}
\begin{center}
\caption{\label{figsharp} Limit of a sharp injector, where again interactions are arbitrary inside the wire. The non-local backscattering Hamiltonian generated by higher order tunneling processes and possible Coulomb interactions become now analogous to an impurity at $x_0$. The results of the paper apply in the presence of initial non-quadratic Hamiltonian $\mathcal{H}_{nq}$, which is taken to be vanishing in the last section such that it does not hide the effective impurity.}
\end{center}
\end{figure}
\begin{figure}[htb]
\vspace{0.3cm}
\begin{center}
\caption{\label{figcoupe} Strong backscattering regime due to the effective impurity at $x_0$ due to invasive effect of the sharp tip. It is reached when  $\omega^*=max[\omega_{12}=eg_L(V_1-V_2), k_B T/\hbar, \omega_L=v/L]$ is less than the crossover energy $\omega_{B}^*$ associated to the backscattering at $x_0$ in the previous figure, and this is even when one has $\omega(x_0)=(V_1+V_2)/2-V(x_0) >\omega_{B}^*$, $V(x_0)$ being the screened potential in Eq.(\ref{V(x)}): the voltage of the tip does not stop the RG flow of $\lambda(x_0)$. Thus for $V_1=V_2$, and in the zero temperature and infinite wire limit, the sharp tip is coupled to a wire disconnected at $x_0$. One could perform perturbative expansion with respect to the tunneling amplitudes $t_{lk}$ where one can expect similar modified scaling laws to those in the weak-backscattering regime in \ref{scaling}.}
\end{center}
\end{figure}
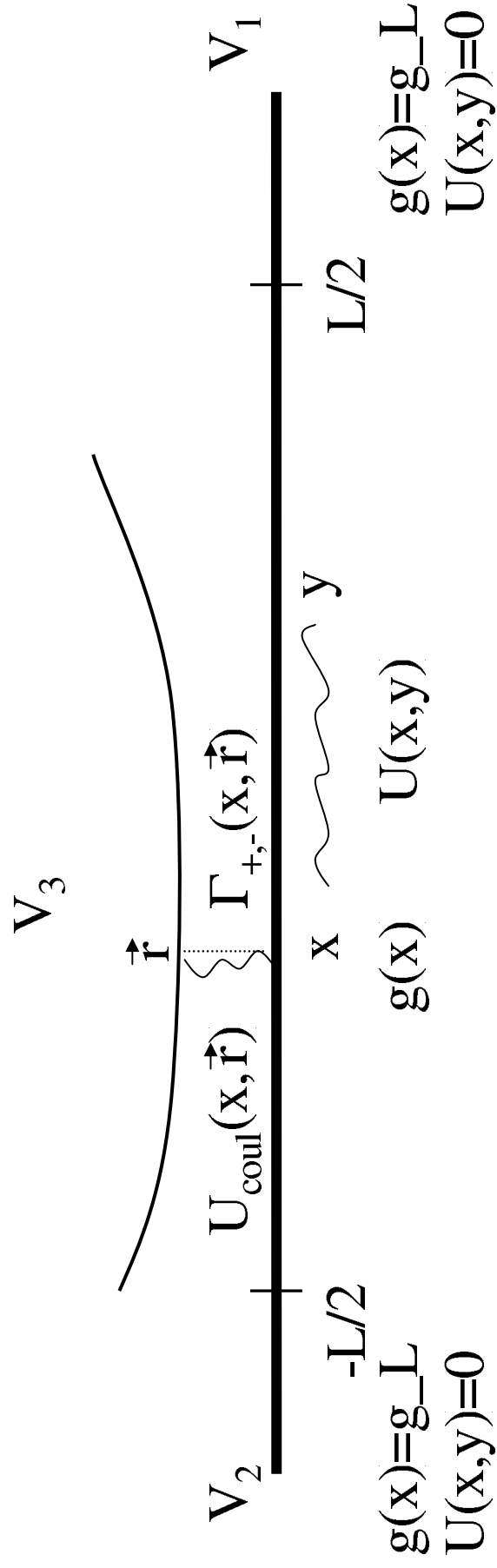
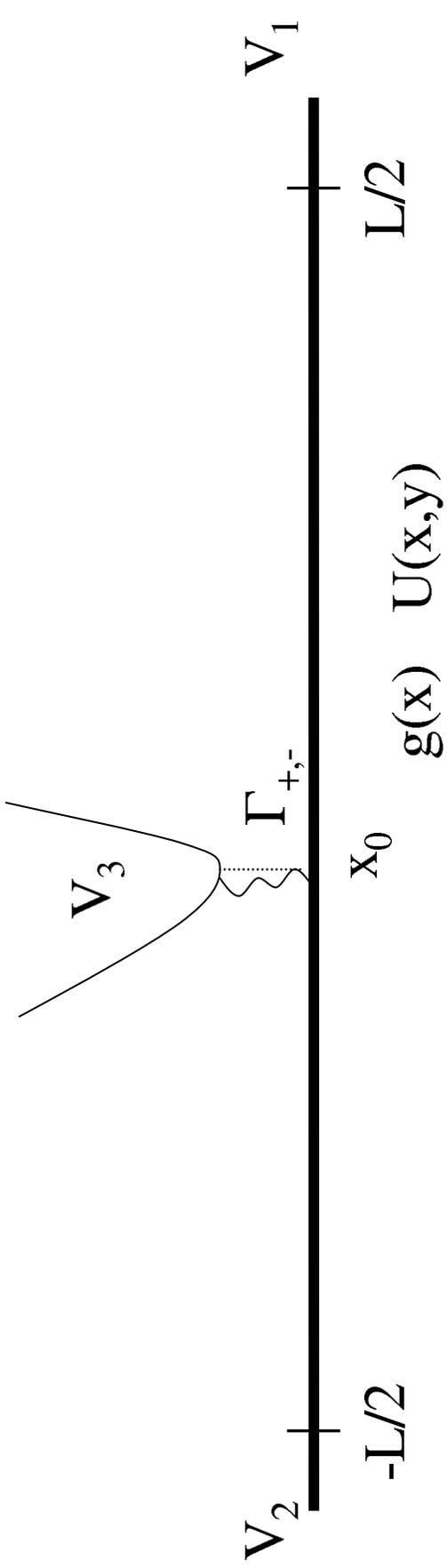
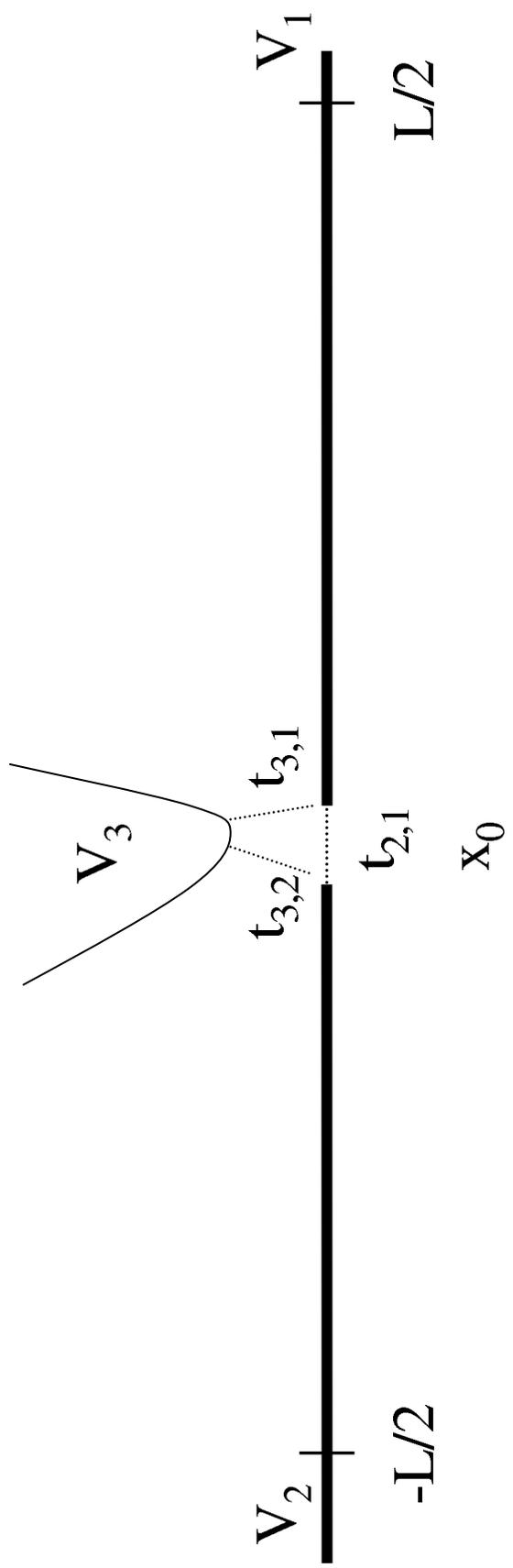
\end{document}